\documentclass[twocolumn]{aastex631}
\usepackage{showyourwork}
\usepackage{multirow}
\usepackage{nicefrac}
\usepackage{wasysym}
\usepackage{amssymb}
\usepackage[inline]{enumitem}
\usepackage{booktabs}

\usepackage[acronym,toc]{glossaries}
\newcommand*\myglsentry[1]{%
  \ifglsused{#1}{%
    \glsentryshort{#1}%
  }{%
    \glsentrylong{#1}%
  }%
}
\newacronym{scc}{SCC}{squamous cell carcinoma}
\newacronym{hnscc}{HNSCC}{head and neck \myglsentry{scc}}
\newacronym{opscc}{OPSCC}{oropharyngeal \myglsentry{scc}}
\newacronym{ctv}{CTV}{clinical target volume}
\newacronym{ctv-n}{CTV-N}{elective \myglsentry{ctv}}
\newacronym{lnl}{LNL}{lymph node level}
\newacronym{hpv}{HPV}{human papilloma virus}
\newacronym{hmm}{HMM}{hidden Markov model}
\newacronym{rv}{RV}{random variable}
\newacronym{dag}{DAG}{directed acyclic graph}
\newacronym{mcmc}{MCMC}{Markov chain Monte Carlo}
\newacronym{ti}{TI}{thermodynamic integration}
\newacronym{bic}{BIC}{Bayesian information criterion}
\newacronym{bn}{BN}{Bayesian network}
\newacronym{ct}{CT}{computed tomography}
\newacronym{mri}{MRI}{magnetic resonance imaging}
\newacronym{pet}{PET}{positron emission tomography}
\newacronym{fna}{FNA}{fine needle aspiration}
\newacronym{clb}{CLB}{Centre Leon Bérard}
\newacronym{usz}{USZ}{University Hospital Zurich}
\newacronym{isb}{ISB}{Inselspital Bern}

\begin{document}

\title{Modelling the Lymphatic Metastatic Progression Pathways of OPSCC \\ from Multi-Institutional Datasets}

\def \USZ {Dep. of Radiation Oncology, University Hospital Zurich, Rämistrasse 100, 8091 Zurich (CH)}
\def \UZH {Dep. of Physics, University of Zurich,  Rämistrasse 71, 8006 Zurich (CH)}

\def \ISBrao {Dep. of Radiation Oncology, Bern University Hospital, University of Bern, Freiburgstrasse 18, 3010, Bern (CH)}
\def \ISBsurgery {Dep. of ENT, Head \& Neck Surgery, Inselspital, Bern University Hospital, University of Bern, Freiburgstrasse 18, 3010 Bern (CH)}
\def \ISBcancer {Head and Neck Anticancer Center, Bern University Hospital, University of Bern, Freiburgstrasse 18, 3010, Bern (CH)}
\def \ISBpath {Institute of Tissue Medicine and Pathology, Bern University Hospital, University of Bern, Murtenstrasse 31, 3008 Bern (CH)}

\def \RHNe {Dep. of ENT, Head \& Neck Surgery, R\'eseau Hospitalier Neuch\^atelois (RHNe),  Neuch\^atel (CH)}

\def \CLBrao {Dep. of Radiation Oncology, Centre L\'eon B\'erard, 28 Rue Laennec, 69008 Lyon (FR)}
\def \CLBsurgery {Dep. of Head and Neck surgery, Centre L\'eon B\'erard, 28 Rue Laennec, 69008 Lyon (FR)}

\def \KS {Institute of Pathology, Klinikum Stuttgart, Kriegsbergstr. 60c, 70174 Stuttgart (GER)}

\author[0000-0001-9434-328X]{Roman Ludwig}
\affiliation{\USZ}
\affiliation{\UZH}

\author{Adrian Schubert}
\affiliation{\RHNe}
\affiliation{\ISBsurgery}
\affiliation{\ISBcancer}

\author{Dorothea Barbatei}
\affiliation{\CLBrao}

\author{Lauence Bauwens}
\affiliation{\CLBrao}

\author[0000-0003-1193-0863]{Jean-Marc Hoffmann}
\affiliation{\USZ}

\author{Sandrine Werlen}
\affiliation{\ISBsurgery}
\affiliation{\ISBcancer}

\author[0000-0002-6996-0646]{Olgun Elicin}
\affiliation{\ISBrao}

\author[0000-0003-0948-1392]{Matthias Dettmer}
\affiliation{\KS}
\affiliation{\ISBpath}

\author[0000-0001-6223-6113]{Philippe Zrounba}
\affiliation{\CLBsurgery}

\author[0000-0002-8060-7827]{Bertrand Pouymayou}
\affiliation{\USZ}

\author[0000-0001-5261-6446]{Panagiotis Balermpas}
\affiliation{\USZ}

\author{Vincent Grégoire}
\affiliation{\CLBrao}

\author[0000-0002-5574-3210]{Roland Giger}
\affiliation{\ISBsurgery}
\affiliation{\ISBcancer}

\author[0000-0002-4275-990X]{Jan Unkelbach}
\affiliation{\USZ}
\affiliation{\UZH}

\correspondingauthor{Roman Ludwig}
\email{roman.ludwig@usz.ch}

\begin{abstract}
    The \gls{ctv-n} in \gls{opscc} is currently based mostly on the prevalence of lymph node metastases in different \glspl{lnl} for a given primary tumor location. We present a probabilistic model for ipsilateral lymphatic spread that can quantify the microscopic nodal involvement risk based on an individual patient's T-category and clinical involvement of \glspl{lnl} at diagnosis.

    We extend a previously published \gls{hmm}, which models the \glspl{lnl} (I, II, III, IV, V, and VII) as hidden binary \glspl{rv}. Each represents a patient's true state of lymphatic involvement. Clinical involvement at diagnosis represents the observed binary \glspl{rv} linked to the true state via sensitivity and specificity. The primary tumor and the hidden \glspl{rv} are connected in a graph. Each edge represents the conditional probability of metastatic spread per abstract time-step, given disease at the edge's starting node. To learn these probabilities, we draw \acrlong{mcmc} samples from the likelihood of a dataset (%
  686\label{output/num_patients.txt}\unskip%
\gls{opscc} patients) from three institutions. We compute the model evidence using \acrlong{ti} for different graphs to determine which describes the data best.

    The graph maximizing the model evidence connects the tumor to each \gls{lnl} and the \glspl{lnl} I through V in order. It predicts the risk of occult disease in level IV is below 5\% if level III is clinically negative, and that the risk of occult disease in level V is below 5\% except for advanced T-category (T3 and T4) patients with clinical involvement of levels II, III, and IV.

    The provided statistical model of nodal involvement in \gls{opscc} patients trained on multi-institutional data may guide the design of clinical trials on volume-deescalated treatment of \gls{opscc} and contribute to more personal guidelines on elective nodal treatment.
\end{abstract}

\section{Introduction}
\label{sec:intro}

\glsunset{ctv}
\glsunset{scc}

When treating \gls{hnscc} with radiotherapy or surgery, the aim is to irradiate or resect as much of the malignant tissue as possible. This includes the primary tumor mass and clinically detected lymph node metastases. However, to reduce the risk of locoregional failure, treatment also includes regions of the lymph drainage system of the neck with possible microscopic tumor spread, which in-vivo imaging modalities such as \gls{ct}, \gls{mri}, or \gls{pet} cannot detect. This is referred to as elective nodal irradiation or prophylactic neck dissection. Treatment decisions regarding the \gls{ctv-n} or the extent of neck dissection must balance the conflicting goals of treating regions at risk of occult lymph node metastases to avoid recurrences while avoiding toxicity related to unnecessary treatment of healthy tissues.

This work concerns itself with \gls{opscc}, where approximately 70-80\% of patients present with lymph node metastases at the time of diagnosis.   In clinical practice, \gls{ctv-n} definition in radiotherapy is based on guidelines \citep{gregoire_ct-based_2003,gregoire_delineation_2014,gregoire_delineation_2018,eisbruch_intensity-modulated_2002,biau_selection_2019,chao_determination_2002,vorwerk_guidelines_2011,ferlito_elective_2009} that are mostly derived from the observed prevalence of involvement in an \gls{lnl} for a given tumor location. These guidelines currently suggest extensive irradiation of both sides of the neck for most patients. In the ipsilateral neck, the \gls{ctv-n} includes \glspl{lnl} II, III and IV for all patients, and levels I and V for the majority of patients. These guidelines, however, do not account for the personal risk of the patients that may depend greatly on their state of tumor progression at diagnosis. E.g., a patient with macroscopic metastases detected via \gls{pet} in both \glspl{lnl} II and III may have a substantial risk for occult disease in \gls{lnl} IV. Instead, patients who present with a clinically N0 neck or a single metastasis in \gls{lnl} II may have a much smaller risk for occult disease in \gls{lnl} IV.

We previously developed a model of lymphatic metastatic progression for estimating the risk of microscopic disease, given a patient's personal diagnosis. The initial model was based on the methodology of \glspl{bn} \citep{pouymayou_bayesian_2019}. It was subsequently extended and formulated as an \gls{hmm} to include T-category in an intuitive manner \citep{ludwig_hidden_2021}. However, these models were introduced based on only a small dataset of approximately 100 early T-category patients available at that time \citep{sanguineti_defining_2009}. The limited data did not allow us to quantify the probability of metastases in the rarely involved \glspl{lnl} I, V, and VII, nor did the data allow us to verify that the \gls{hmm} is adequate to describe the dependence on lymph node involvement on T-category. In this paper, we extend the previous work \citep{ludwig_hidden_2021} by making the following contributions:
\begin{enumerate}
    \item We provide a \gls{hmm} of ipsilateral lymph node involvement including all relevant \glspl{lnl}, namely the levels I, II, III, IV, V, and VII. To determine the optimal underlying \gls{dag} we compare different graphs by calculating the model evidence through \gls{ti}.
    \item We collect a multi-centric dataset consisting of %
  686\label{output/num_patients.txt}\unskip%
patients from three institutions, allowing us to train the model based on a sizable dataset \citep{ludwig_dataset_2022,ludwig_multi-centric_2023}.
    \item We use the trained model to provide personalized risk estimations for occult metastases for typical clinical states of tumor progression at diagnosis, illustrating its potential for guiding volume-deescalated treatment strategies in the future. 
\end{enumerate}

\section{HMM Formalism and Notation}
\label{sec:formalism}

\glsreset{bn}

\begin{figure}
    \centering
    \includegraphics[width=0.6\linewidth]{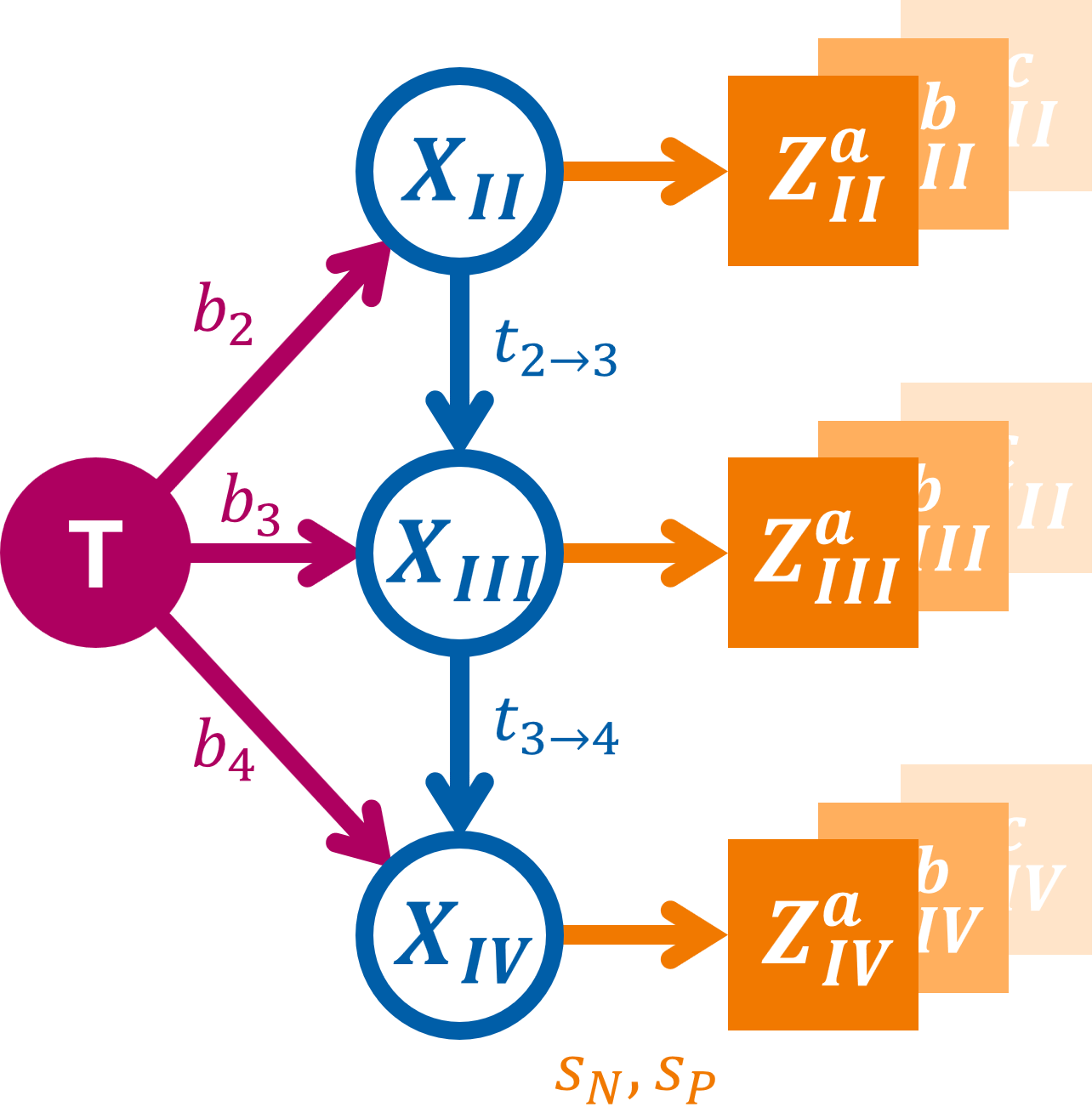}
    \caption{\Gls{dag} representing a possible abstraction of the lymphatic network comprising the tumor (red shaded circle) and \glspl{lnl} II through IV as hidden binary \glspl{rv} (blue outlined circles). Attached to each of these is the corresponding observed \gls{rv} (orange shaded squares). Lymphatic flow is depicted in the form of parameterized arrows (red and blue) that represent the probability of spread along the respective arc per time-step. Sensitivity and specificity (orange arrows) connect the hidden \glspl{rv} to the diagnosis. \label{fig:graph_with_obs}}
\end{figure}

\subsection{State of the Hidden Markov Model}
\label{subsec:formalism:state}

We have introduced a probabilistic model for lymph node involvement based on \glspl{bn} in \citep{pouymayou_bayesian_2019}. The model was extended using \glspl{hmm} in \citep{ludwig_hidden_2021}. We will briefly recap the \acrlong{hmm} to introduce the notation used throughout the work.

A patient's state of (hidden) lymphatic involvement at time $t$ is described as a collection of binary \glspl{rv}, one for each of the $V$ \glspl{lnl}:
\begin{equation}
    \mathbf{X}[t] = \left( X_v[t] \right) \qquad v \in \left\{ 1,2, \ldots, V \right\}
\end{equation}
Each of the \glspl{lnl} can be in the state $X_v=0$ (\texttt{FALSE}), meaning \gls{lnl} $v$ is healthy, or in the state $X_v=1$ (\texttt{TRUE}), indicating the \gls{lnl} harbors metastases. The involved state includes occult disease.

The transition from one time-step to another is governed by the transition probability $P\left( \mathbf{X}[t+1]=\boldsymbol{\xi}_i \mid \mathbf{X}[t]=\boldsymbol{\xi}_j \right)$, which can conveniently be collected into a transition matrix when we enumerate all $2^V$ distinct possible states $\boldsymbol{\xi}_i$ with $i \in \left\{ 1,2, \ldots, 2^V \right\}$ of lymphatic involvement:
\begin{equation}
    \mathbf{A} = \left( A_{ij} \right) = \left( P\left( \mathbf{X}[t+1]=\boldsymbol{\xi}_i \mid \mathbf{X}[t]=\boldsymbol{\xi}_j \right) \right)
\end{equation}
The term $P\left( \boldsymbol{\xi}_i \mid \boldsymbol{\xi}_j \right)$ describes the probability to transition from the hidden state of lymphatic involvement $\boldsymbol{\xi}_j$ to the state $\boldsymbol{\xi}_i$ between the time $t$ and $t+1$. Using a \gls{dag} as depicted in \autoref{fig:graph_with_obs}, we can formulate this transition probability in the following way:
\begin{equation}
    \label{eq:transition_prob}
    P\left( \boldsymbol{\xi}_i \mid \boldsymbol{\xi}_j \right) = \prod_{v \leq V} Q\left( \xi_{iv} ; \xi_{jv} \right) P \left( \xi_{iv} \mid \left\{ \xi_{jr} \right\}_{r \in \operatorname{pa}(v)} \right)^{1 - \xi_{jv}}
\end{equation}
In this equation, we have denoted \glspl{lnl} that are parents of \gls{lnl} $v$ with the symbol $r\in\operatorname{pa}(v)$. Also, $\xi_{iv}$ denotes the value that \gls{lnl} $v$ takes on when the patient is in state $\boldsymbol{\xi}_i$. The term $Q(a;b) \in \{ 0,1 \}$ is there to prohibit self-healing. It is always one, except if \gls{lnl} $v$ is healthy in state $\xi_i$, but was metastatic in the previous state $\xi_j$. In that case the function becomes $Q(0;1) = 0$, making the transition back to healthier states impossible.

The terms of the form $P \left( \xi_{iv} \mid \left\{ \xi_{jr} \right\}_{r \in \operatorname{pa}(v)} \right)$ implicitly depend on how we parameterize the arcs of \autoref{fig:graph_with_obs}. For example, if we look at the probability of spread to \gls{lnl} III ($X_3$) depending on the state of that level's parent -- which, in this case, is $\operatorname{pa}(3) = 2$ -- we can write the different combinations into a conditional probability table as below.

\begin{deluxetable}{cc|cc}
\tablecaption{$P\left( X_3 \mid X_2 \right)$ for all possible combinations of $X_3$ and $X_2$.}
\tablehead{& & \colhead{$X_2=0$} & \colhead{$X_2=1$}}
\startdata
\begin{tabular}{@{}c@{}}
\multirow{2}{*}{$X_3$}
\end{tabular} & $=0$ & $1-b_3$ & $(1-b_3)(1-t_{23})$ \\
& $=1$ & $b_3$ & $1-b_3-t_{23}+b_3 t_{23}$ \\
\enddata
\end{deluxetable}

The variable $b_3$ denotes the probability of lymphatic spread from the tumor to \gls{lnl} III during one time-step, and $t_{23}$ is the probability of spread from an involved level II further down the lymphatic chain into \gls{lnl} III. 

\subsection{Diagnostic Observation}
\label{subsec:formalism:diagnosis}

We also need to introduce a separate collection of \glspl{rv} that describe the diagnostic observation of a patient's involvement. In analogy to the hidden true state $\mathbf{X}[t]$ at time $t$, we write this diagnosis as
\begin{equation}
    \mathbf{Z} = \left( Z_v \right) \qquad v \in \left\{ 1,2, \ldots, V \right\}
\end{equation}
We do not need to differentiate between different times $t$ here, since a patient is ever only diagnosed once, after which treatment usually starts timely.
Diagnosis and true state of a patient are formally connected via the sensitivity $s_N$ and specificity $s_P$ of the used diagnostic modality. In clinical practice, these modalities are \gls{ct}, \gls{mri}, or \gls{pet} scan, but it may also include information from biopsies after a \gls{fna} or other techniques to detect lymphatic metastases. For each \gls{lnl} $v$ the conditional probability table of $P\left( Z_v \mid X_v \right)$ looks like this:

\begin{deluxetable}{cc|cc}
\tablecaption{$P\left( Z \mid X \right)$ for all possible values of $X$ and $Z$.}
\tablehead{& & \colhead{$X=0$} & \colhead{$X=1$}}
\startdata
\begin{tabular}{@{}c@{}}
\multirow{2}{*}{$Z$}
\end{tabular} & $=0$ & $s_P$ & $1-s_N$ \\
& $=1$ & $1-s_P$ & $s_N$ \\
\enddata
\end{deluxetable}

Consequently, the conditional probability to observe a diagnosis $\mathbf{Z}=\boldsymbol{\zeta}_\ell$, given a hidden involvement state $\mathbf{X}=\boldsymbol{\xi}_k$ is a matrix $\mathbf{B}$ made up of products of terms from the table above:
\begin{equation} \label{eq:transition_matrix}
    \mathbf{B} = \left( B_{k\ell} \right) = \prod_{v=1}^V P\left( Z_v = \zeta_{\ell v} \mid X_v[t_\text{D}] = \xi_{kv} \right)
\end{equation}

We define the time $t=0$ to be the moment just before a patient's tumor formed, and hence $X_v[t=0]=0 \,\,\, \forall v$. However, using this definition, we cannot know how many time-steps have passed until $t_\text{D}$, when the patient was diagnosed with cancer. We can only make the assumption that a patients with an earlier T-category tumor was \emph{probably} diagnosed after fewer time-steps than a patient with a advanced T-category tumor.

We can use this assumption by marginalizing over the diagnose times $t_\text{D}$ of patients in different T-categories using different prior distributions over the diagnose time. E.g., $P \left( t=t_\text{D} \mid \text{early} \right)$ for early T-category patients (T1 \& T2) and $P \left( t=t_\text{D} \mid \text{late} \right)$ for advanced T-category patients (T3 \& T4). Throughout this work we will use binomial distributions for these probability mass functions, mainly because they have a plausible shape for this purpose and only a single parameter.
\begin{equation} \label{eq:time_dist}
    P \left( t = t_\text{D} \mid \text{T}x \right) = \operatorname{\mathfrak{B}}(t_\text{max},p_{\text{T}x})
\end{equation}
Here, the parameter $p_{\text{T}x}$ can be interpreted as the probability that the patient with a tumor of T-category $x$ will be diagnosed at time-step $t+1$ given they are in time-step $t$. We will use as the latest time-step $t_\text{max} = 10$, which will therefore give us a distribution over the diagnosis time that has its mean at $\mathbb{E}[t_\text{D}] = 10 \cdot p_{\text{T}x}$.

\subsection{The Likelihood Function}
\label{subsec:formalism:likelihood}

Using the definitions up to this point, we can compute a vector of likelihoods for every possible diagnosis:
\begin{equation} \label{eq:likelihood_vec}
\begin{split}
    \boldsymbol{\ell} &= \big( P\left( \mathbf{Z} = \boldsymbol{\zeta}_i \right) \big) \\
    &= \sum_{t=0}^{t_{max}} \left[ \boldsymbol{\pi} \cdot \mathbf{A}^t \cdot \mathbf{B} \right] \cdot P \left( t \mid \text{T} \right)
\end{split}
\end{equation}
This likelihood implicitly depends on how we parameterize the arcs of the \gls{dag} underlying the model -- see \autoref{eq:transition_prob} -- and the parameterization of the distribution over diagnosis times -- e.g., as in \autoref{eq:time_dist}. 

Together with the parametrizations of the distributions over the diagnosis time, the parameters $b_v$ and $t_{vr}$ that make up the transition matrix $\mathbf{A}$ comprises the set of model parameters:
\begin{equation}
    \boldsymbol{\theta} = \left( \left\{ b_v \right\}, \left\{ t_{vr} \right\}, p_\text{early}, p_\text{late} \right) \quad \text{with} \quad \genfrac{}{}{0pt}{2}{v\leq V}{r\in\operatorname{pa}(v)}
\end{equation}

To infer these parameters from a dataset of $N$ \gls{opscc} patients $\boldsymbol{\mathcal{D}} = \left( d_1, d_2, \ldots, d_N \right)$, we compute the data log-likelihood:
\begin{equation} \label{eq:log_likelihood}
    \log\mathcal{L} \left( \boldsymbol{\mathcal{D}} \mid \boldsymbol{\theta} \right) = \sum_{i=1}^N \log P \left( \mathbf{Z} = d_i \right)
\end{equation}
Which effectively amounts to computing the element-wise logarithm of the likelihood vector $\boldsymbol{\ell}$ from \autoref{eq:likelihood_vec} and summing up the entries that correspond to each of the patients $d_i$ for $i\leq N$.

Note that it is also possible to account for incomplete diagnoses, i.e., a diagnosis where the involvement information for one or more \glspl{lnl} is missing. In that case, we can sum over those elements of $\boldsymbol{\ell}$ that correspond to complete diagnoses which match the provided incomplete one. In this paper, for some patients involvement information of level VII was missing and hence marginalized over. A detailed explanation of this formalism can be found in \citep{zora231470}, section 6.2.7.

Using this log-likelihood function one may now employ a variety of inference methods to learn the parameters of the model that best describe the observed data.

\subsection{Parameter Inference}
\label{subsec:formalism:inference}

\begin{figure}
    \centering
    \includegraphics[width=0.6\linewidth]{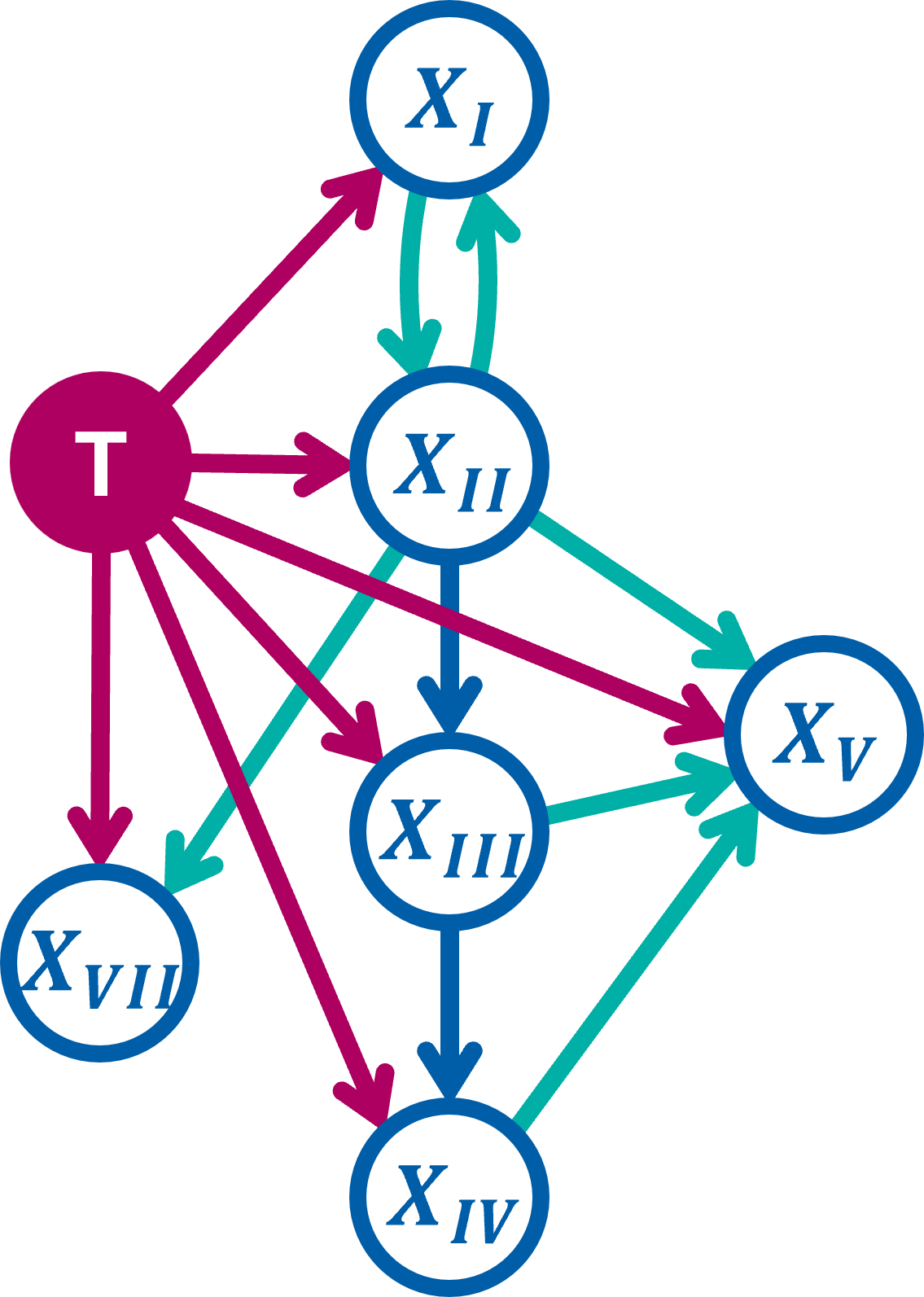}
    \caption{Extended \gls{dag} representing different possible spread graphs underlying the \gls{hmm}. As in \autoref{fig:graph_with_obs}, red arcs are parametrized with probabilities of spread from the tumor (red circle) to the \glspl{lnl} (blue circles). These red arcs, together with the blue arcs fron \gls{lnl} to \gls{lnl}, make up the \emph{base graph}. One after the other, each of the green arcs was added to the base graph. Subsequently, the performance of the resulting models in terms of its \gls{bic} was compared to the base graph to assess whether the additional edge should be kept in the \emph{winning graph} or not. \label{fig:extended_graph}}
\end{figure}

We use \gls{mcmc} sampling to draw parameter samples $\boldsymbol{\hat{\theta}}_i$ for $i \leq S$ from the likelihood described in \autoref{eq:log_likelihood} (i.e. the unnormalized posterior distribution over the parameters $\boldsymbol{\theta}$, since we used a uniform prior in this work).

More specifically, we use the Python implementation \texttt{emcee} \citep{foreman-mackey_emcee_2013} and two sample proposal mechanisms based on differential evolution moves \citep{ter_braak_differential_2008,nelson_run_2013} for sampling. Instead of proposing and then accepting or rejecting individual parameter samples one after the other (as in the classical Metropolis-Hastings algorithm), the \texttt{emcee} implementation makes use of an ensemble of $W$ so-called ``walkers''. This gives rise to $W$ parallel chains of samples that mutually influence each others proposal such that the sampling proceedure overall is \emph{affine invariant}. This means that scaling the parameter space along any dimension has no effect on the performance of the \gls{mcmc} sampling algorithm.

For the experiments in this work, we used $W = 20 \cdot k$ walkers, where $k$ is the dimensionality of the parameter space $\boldsymbol{\Theta}$. After an initial ``burn-in'' phase, during which all drawn samples are discarded because they are not yet independent of the initial state, we continued sampling for another 200 steps of which we discarded every 10th to be left with $S = 20 \cdot W$ samples.

These $S$ parameter estimates are then used to compute expectation values of estimates that depend on the parameters $\boldsymbol{\theta}$ through an integral over the parameter space $\boldsymbol{\Theta}$:

\begin{equation}
    \begin{aligned}
        \mathbb{E}_p \left[ f \right] &= \int_{\boldsymbol{\Theta}} p(\boldsymbol{\theta}) f(\boldsymbol{\theta}) d\boldsymbol{\theta} \\
        &\approx \frac{1}{S} \sum_{i=1}^S f \big( \boldsymbol{\hat{\theta}}_i \big)
    \end{aligned}
\end{equation}
Alternatively, the individual $\hat{f}_i = f\big( \boldsymbol{\hat{\theta}}_i \big)$ can be used to plot histograms over the distribution of $f$. We will do so in \autoref{sec:results} to show distributions over prevalence predictions and risk computations.

Another relevant model parameter that needs to be set for the inference process, is the maximum number of time-steps we used for the evolution of the system. We set this value to $t_\text{max} = 10$, such that $t \in \{ 0, 1, 2, \ldots, 10 \}$. The binomial ``success probability'' used to fix the shape of the early T-category's time-prior was set to $p_\text{early} = 0.3$.

\begin{figure}[t]
    \script{thermo_int.py}
    \centering
    \includegraphics[width=\linewidth]{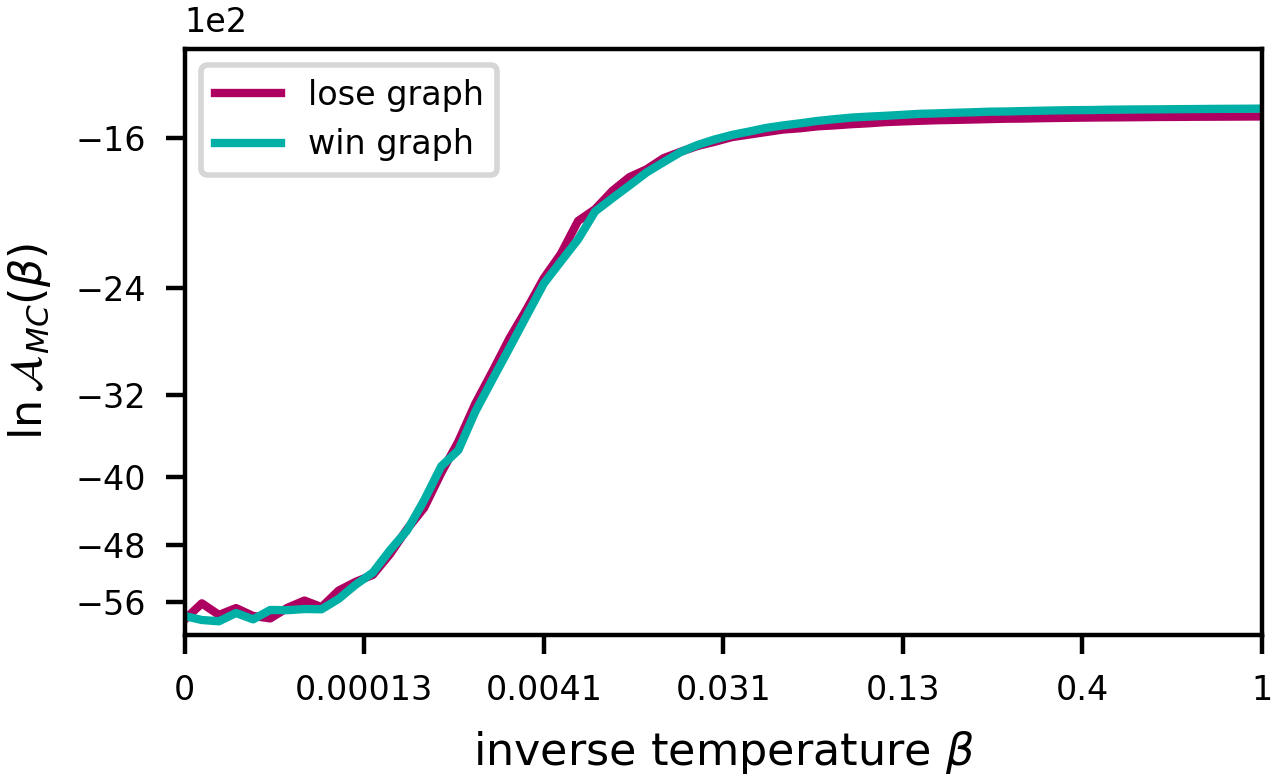}
    \caption{The expected log-accuracy under the power posterior plotted against the value of the inverse temperature $\beta$. These values were computed during a \gls{ti} run for the loosing graph model (red line), as well as the winning graph model (green line). Note that the scaling of the x-axis is chosen such that the 64 points on the temperature ladder appear evenly spaced. This is to stress how the log-accuracy develops in the range from $\beta=0$ to around $\beta=0.1$. The area under these two curves yields an approximation of the log-evidence of the respective model. It is barely visible that due to the simplicity of the loosing graph model its accuracy increases earlier and faster that the winning graph model's. However, already before $\beta=0.031$ the model based on the winning graph outperforms the loosing graph model. Ultimately, the winning graph's log-evidence is better than the loosing graph model's by a value of around %
  \input{output/evidence_diff.txt}\label{output/evidence_diff.txt}\unskip%
(see \autoref{table:evidence}). \label{fig:thermo_int}}
\end{figure}

\subsection{Risk estimation}
\label{subsec:formalism:risk}

The main task for personalizing the \gls{ctv-n} definition is to predict the probability of the hidden possible states $\boldsymbol{\xi}_k$ given the diagnosis $d^\star=\boldsymbol{\zeta}_\ell$ of a new patient. Using Bayes' theorem, we get

\begin{equation}
    P\left( \mathbf{X}=\boldsymbol{\xi}_k \mid \mathbf{Z}=\boldsymbol{\zeta}_\ell \right) = \frac{P\left( \boldsymbol{\zeta}_\ell \mid \boldsymbol{\xi}_k \right) P\left( \boldsymbol{\xi}_k \mid \boldsymbol{\theta} \right)}{\sum_{r=1}^{2^V} P\left( \boldsymbol{\zeta}_\ell \mid \boldsymbol{\xi}_r \right) P\left( \boldsymbol{\xi}_r \mid \boldsymbol{\theta} \right) }
\end{equation}
The described model along with the inferred parameters $\boldsymbol{\hat{\theta}}$ will yield an estimate (or multiple estimates) for the ``prior'' in the above equation $P\big( \boldsymbol{\xi}_k \mid \boldsymbol{\hat{\theta}} \big)$.

From this probability for any possible hidden state, we can also compute the probability of, for example, involvement in \gls{lnl} IV. To that end, we marginalize over all states $\boldsymbol{\xi}_k$ where $\xi_{k4} = 1$, meaning those states in which \gls{lnl} IV harbors metastases. Formally, we can define a marginalization vector $\mathbf{m}$ that is one for every hidden state we want to include in the marginalization and zero elsewhere. In the example of the marginalized probability for \gls{lnl} IV involvement, the components would look like this:
\begin{equation}
    m_{4k} = \operatorname{id}(\xi_{k4} = 1)
\end{equation}
Subsequently, we can compute the marginalization as a dot product:
\begin{equation}
    \begin{aligned}
        P\left(\text{IV} = 1 \mid \mathbf{Z}=\boldsymbol{\zeta}_\ell \right) &= \sum_{k:\xi_{k4} = 1} P\left( \mathbf{X}=\boldsymbol{\xi}_k \mid \mathbf{Z}=\boldsymbol{\zeta}_\ell \right) \\
        &= \mathbf{m}_4 \cdot P\left( \mathbf{X}=\boldsymbol{\xi}_k \mid \mathbf{Z}=\boldsymbol{\zeta}_\ell \right)
    \end{aligned}
\end{equation}

\section{Complete Model of Ipsilateral Spread in OPSCC}
\label{sec:complete_model}

\subsection{Investigating Spread Graphs}
\label{subsec:complete_model:graphs}

The \gls{dag} shown in \autoref{fig:graph_with_obs} includes the \glspl{lnl} II, III, and IV, which represent the most relevant lymph node levels for \gls{opscc}. It includes the arcs from II to III and from III to IV, representing the main direction of lymphatic drainage, which is well motivated anatomically and by the data on lymph node involvement. Previous publications  \citep{pouymayou_bayesian_2019,ludwig_hidden_2021} focused on these levels because they relied on a limited reconstructed dataset of \gls{opscc} patients \citep{sanguineti_defining_2009}. Now, with the datasets available for this work, we can extend the graph to include random variables for all \glspl{lnl} that are relevant for \gls{opscc}: I, II, III, IV, V, and VII. The main question to answer is: which arcs between \glspl{lnl} are needed to accurately model the data on lymph node involvement without increasing the model's complexity unnecessarily.

First, we notice that the direct arcs from tumor to each of the \glspl{lnl} must be present, since every \gls{lnl} appears metastatic in isolation at least once in the dataset. For example, some patients presented with metastases in \gls{lnl} I, while the other levels appeared healthy. If no spread was allowed from the tumor to $X_1$ (i.e., $b_1 = 0$), the likelihood of observing this patient would be zero.

We have more freedom in choosing how to connect the \glspl{lnl} to each other. To investigate which connections to add we start by establishing a baseline from a model using a minimal \emph{base graph}. It contains only the connections from \gls{lnl} II to III and an arc from \gls{lnl} III to IV, as motivated by the \emph{main lymphatic pathway} \citep{lengele_anatomical_2007}. The base graph is illustrated in \autoref{fig:extended_graph} via the red and blue arcs. The lymphatic drainage to or from levels I, V, and VII is not as clearly defined. Therefore, we define a set of candidate arcs (green arcs in \autoref{fig:extended_graph}) and use the model comparison methodology described in \autoref{subsec:complete_model:comparison} to determine which graph is most supported our data. 

To connect level I, we investigate two candidate arcs: from I to II and from II to I. An arc from I to II was used in \citep{pouymayou_bayesian_2019,ludwig_hidden_2021} and is anatomically motivated. However, since \gls{lnl} I is rarely involved compared to level II, the associated parameter $t_{12}$ is mostly undetermined. Therefore, we also consider the flipped arc from \gls{lnl} II to I and investigate if it helps to describe the correlations between the involvement of levels I and II.

Anatomically, the \emph{posterior accessory pathway} that drains \gls{lnl} II through \gls{lnl} V motivates investigating and arc from $X_2$ into $X_5$ \citep{lengele_anatomical_2007}. And, although no lymphatic pathway is described that directly drains \gls{lnl} III or IV into level V, due to their proximity to each other, we will also investigate additional edges from the levels III and IV into \gls{lnl} V. Finally, we look at adding an arc from \gls{lnl} II to \gls{lnl} VII also due to their anatomical proximity.

To determine the optimal graph, we first consider six models, each with one of the six green candidate arcs of \autoref{fig:extended_graph} added to the base graph. Every one of these models was evaluated by computing an approximation to its evidence via thermodynamic integration as described below in \autoref{subsec:complete_model:comparison}. Subsequently, graphs combining multiple arcs that individually improve the model evidence are considered. Thereby, the ``winning graph'' is determined, which yields the highest (i.e., the least negative) value of the logarithm of the model evidence.\looseness=-1

\subsection{Model Comparison}
\label{subsec:complete_model:comparison}

\begin{deluxetable}{c|c|c}
\tablecaption{
    Interpretation of Bayes factors and their natural logarithms in terms of their support for or against one of the two compared models as introduced by \citet{jeffreys_theory_1998}.%
    \label{table:bayes_factor}%
}
\tablehead{
    \colhead{$K_\text{1v2}$} & \colhead{$\ln{K_\text{1v2}}$} & \colhead{support for $\mathcal{M}_1$}
}
    \startdata
    $< 10^0$ & $< 0$ & negative (supports $\mathcal{M}_2$) \\
    $10^0$ to $10^{\nicefrac{1}{2}}$ & 0 to 1.15 & barely worth a mention \\
    $10^{\nicefrac{1}{2}}$ to $10^1$ & 1.15 to 2.3 & substantial \\
    $10^1$ to $10^{\nicefrac{3}{2}}$ & 2.3 to 3.45 & strong \\
    $10^{\nicefrac{3}{2}}$ to $10^2$ & 3.45 to 4.6 & very strong \\
    $> 10^2$ & $> 4.6$ & decisive \\
    \enddata
\end{deluxetable}

The aim of this work is to refine the graph structure underlying our risk model introduced in the previous section. This \gls{dag} determines the number of parameters of the model as well as how exactly the transition matrix $\mathbf{A}$ is parameterized. To compare different models that are based on different \glspl{dag}, e.g. models $\mathcal{M}_1$ and $\mathcal{M}_2$, in a Bayesian setting, we need to compute the probabilities of these models, given the data $\boldsymbol{\mathcal{D}}$:
\begin{equation}
    P \left( \mathcal{M}_i \mid \boldsymbol{\mathcal{D}} \right) = \frac{P \left( \boldsymbol{\mathcal{D}} \mid \mathcal{M}_i \right) P \left( \mathcal{M}_i \right)}{ P \left( \boldsymbol{\mathcal{D}} \right) }
\end{equation}
If we assume all models $\mathcal{M}_i$ for $i \in \{ 1,2 \}$ to have the same \emph{a priori} probability -- meaning in this case $P (\mathcal{M}_1) = P (\mathcal{M}_2)$ -- then we can compute the so-called \emph{Bayes factor} of the two models as the ratio of their likelihoods. The interpretation of the values for different Bayes factors is given in \autoref{table:bayes_factor}. It is defined as follows:
\begin{equation}
    K_\text{1v2} = \frac{P \left( \mathcal{M}_1 \mid \boldsymbol{\mathcal{D}} \right)}{P \left( \mathcal{M}_2 \mid \boldsymbol{\mathcal{D}} \right)} = \frac{P \left( \boldsymbol{\mathcal{D}} \mid \mathcal{M}_1 \right)}{P \left( \boldsymbol{\mathcal{D}} \mid \mathcal{M}_2 \right)}
\end{equation}
These likelihoods are commonly called the \emph{model evidence} or \emph{marginal likelihood}. The latter because computing it involves marginalizing the data likelihood over all model parameters:
\begin{equation} \label{eq:evidence}
    E_\mathcal{M} = P \left( \boldsymbol{\mathcal{D}} \mid \mathcal{M} \right) = \int_{\boldsymbol{\Theta}} P\left( \boldsymbol{\mathcal{D}} \mid \boldsymbol{\theta}, \mathcal{M} \right) p(\boldsymbol{\theta} \mid \mathcal{M}) d\boldsymbol{\theta}
\end{equation}
However, this quantity is often very hard to compute or even intractable, due to the high dimensionality of the parameter space $\boldsymbol{\Theta}$. In our case, the number of dimensions ranges from $k=9$ for the \emph{base graph} to $k=11$ for the \emph{winning graph}. A brute-force integration over a unit cube with this many dimensions is inefficient and error-prone, which is why we resorted to \gls{ti} for computing the (log-)evidence.

Below, we will briefly outline the main concept behind this algorithm. An intuitive and extensive derivation of \gls{ti} is given by \citep{aponte_introduction_2022}.

We start by taking the logarithm of the model evidence $E$ and subtract a zero from it in the form of the term $0 = \ln \int p(\boldsymbol{\theta} \mid \mathcal{M}) d\boldsymbol{\theta}$. Further, we can multiply the distribution over the parameters $\boldsymbol{\theta}$ inside this integral by $1 = P \left( \boldsymbol{\mathcal{D}} \mid \boldsymbol{\theta}, \mathcal{M} \right)^{\beta=0}$. Subsequently, we can write the logarithm of the evidence as an integral over a derivative:

\begin{equation} \label{eq:ti}
    \begin{aligned}
        \ln E &= \ln \int P \left( \boldsymbol{\mathcal{D}} \mid \boldsymbol{\theta}, \mathcal{M} \right)^{\beta=1} p \left( \boldsymbol{\theta} \mid \mathcal{M} \right) d\boldsymbol{\theta} - \ln E_0 \\
        &= \int_0^1 \frac{d}{d\beta} \ln E_\beta d\beta
    \end{aligned}
\end{equation}
Where we have used the (unnormalized) \emph{power posterior} $p_\beta \left( \boldsymbol{\theta} \mid \boldsymbol{\mathcal{D}}, \mathcal{M} \right) = P \left( \boldsymbol{\mathcal{D}} \mid \boldsymbol{\theta}, \mathcal{M} \right)^\beta p \left( \boldsymbol{\theta} \mid \mathcal{M} \right)$ to compute the respective evidence $E_\beta = \int p_\beta \left( \boldsymbol{\theta} \mid \boldsymbol{\mathcal{D}}, \mathcal{M} \right) d\boldsymbol{\theta}$.

\begin{deluxetable}{l|rr}
    \tablecaption{
        Literature sensitivity and specificity values that we used to infer the most likely involvement for a patient when multiple diagnostic modalities reported conflicting nodal involvement \citep{de_bondt_Detection_2007,kyzas_18f-fluorodeoxyglucose_2008}.
        \label{table:sens_spec}
    }
    \tablehead{
        \textbf{Modality} & \textbf{Specificity} & \textbf{Sensitivity}
    }
    \startdata
        \acrshort{ct} & 76\% & 81\% \\
        \acrshort{pet} & 86\% & 79\% \\
        \gls{mri} & 63\% & 81\% \\
        \acrshort{fna} & 98\% & 80\% \\
        Pathology & $\approx$ 100\% & $\approx$ 100\% \\
    \enddata
\end{deluxetable}

The derivatives of the log-evidences $\ln E_\beta$ are essentially expectation values of the data log-likelihood under the power posteriors of the corresponding value for $\beta$. They can be computed using \gls{mcmc}:
\begin{equation}
    \begin{aligned}
        \frac{d}{d\beta} \ln E_\beta &= \int p_\beta \left( \boldsymbol{\theta} \mid \boldsymbol{\mathcal{D}}, \mathcal{M} \right) \ln P \left( \boldsymbol{\mathcal{D}} \mid \boldsymbol{\theta}, \mathcal{M} \right) d\boldsymbol{\theta} \\
        &= \mathbb{E} \left[ \ln P \left( \boldsymbol{\mathcal{D}} \mid \boldsymbol{\theta}, \mathcal{M} \right) \right]_{ p_\beta \left( \boldsymbol{\theta} \mid \boldsymbol{\mathcal{D}}, \mathcal{M} \right) } \\
        &\approx \frac{1}{S} \sum_{i=1}^S \ln P \left( \boldsymbol{D} \mid \hat{\boldsymbol{\theta}}_{\beta i}, \mathcal{M} \right) =: \mathcal{A}_\text{MC} \left( \beta \right)
    \end{aligned}
\end{equation}
The integral in \autoref{eq:ti} can then be computed via a trapezoidal rule using the $\mathcal{A}_\text{MC}$ to yield a  numerical approximation of the model evidence:
\begin{equation}
    \ln E \approx \frac{1}{2} \sum_{j=0}^{R-1} \left( \beta_{j+1} - \beta_j \right) \cdot \big( \mathcal{A}_\text{MC} (\beta_{j+1}) + \mathcal{A}_\text{MC} (\beta_j) \big)
\end{equation}
This estimate gets better for more samples $S$ per sampling from the power posterior $p_\beta$ but more importantly it gets better for a tighter spacing of the values for $\beta$ within the interval $[0,1]$. The variable $\beta$ is also often referred to as an \emph{inverse temperature}, due to its origins in statistical physics. Often when performing \gls{ti}, the most drastic changes in the values of the $\mathcal{A}_\text{MC}$ occur at high temperatures (meaning $\beta$ very close to zero), while the changes become smaller and smaller for lower temperatures ($\beta$ towards one). It is therefore efficient to space the \emph{temperature ladder} unevenly, e.g. according to a fifth order power rule:
\begin{equation} \label{eq:power_rule}
    \beta_j = \left( j / R \right)^5 \qquad j \in \{ 0, 1, 2, \ldots, R \}
\end{equation}
For the \glspl{ti} that were performed in this work we used such a fifth order power rule with 64 steps, meaning that $R=63$.

\begin{deluxetable}{ccccccrrrr}
    \tablecaption{Prevalence of involvement patterns in the multi-centric dataset. An involvement pattern is characterized by the state of the six \glspl{lnl}: A red dot means the \gls{lnl} was reported to be metastatic, a green dot means it was determined to be healthy and a question mark means that the prevalence was marginalized over the state of this \gls{lnl}. \label{table:data_prevalence}}
    \tablehead{
        \multicolumn{6}{c}{\textbf{LNL involvement} } & \multicolumn{4}{c}{\textbf{T-category} } \\
        I & II & III & IV & V & VII & \multicolumn{2}{c}{\textbf{early} } & \multicolumn{2}{c}{\textbf{advanced} }
    }
    \startdata
  ? & {\color{red} \CIRCLE} & ? & ? & ? & {\color{red} \CIRCLE} & 10 & (2\%) & 20 & (7\%) \\
? & {\color{green} \CIRCLE} & ? & ? & ? & {\color{red} \CIRCLE} & 4 & (0\%) & 2 & (0\%) \\
? & ? & {\color{red} \CIRCLE} & ? & {\color{red} \CIRCLE} & ? & 12 & (2\%) & 17 & (6\%) \\
? & ? & {\color{green} \CIRCLE} & ? & {\color{red} \CIRCLE} & ? & 16 & (3\%) & 9 & (3\%) \\
? & {\color{red} \CIRCLE} & ? & ? & ? & ? & 305 & (72\%) & 202 & (76\%) \\
? & {\color{red} \CIRCLE} & {\color{red} \CIRCLE} & ? & ? & ? & 100 & (23\%) & 87 & (33\%) \\
? & {\color{red} \CIRCLE} & {\color{green} \CIRCLE} & ? & ? & ? & 205 & (48\%) & 115 & (43\%) \\
? & {\color{green} \CIRCLE} & {\color{red} \CIRCLE} & ? & ? & ? & 18 & (4\%) & 12 & (4\%) \\
? & ? & {\color{red} \CIRCLE} & ? & ? & ? & 118 & (27\%) & 99 & (37\%) \\
? & ? & {\color{red} \CIRCLE} & {\color{red} \CIRCLE} & ? & ? & 25 & (5\%) & 23 & (8\%) \\
? & ? & {\color{red} \CIRCLE} & {\color{green} \CIRCLE} & ? & ? & 93 & (21\%) & 76 & (28\%) \\
? & ? & {\color{green} \CIRCLE} & {\color{red} \CIRCLE} & ? & ? & 6 & (1\%) & 8 & (3\%) \\
? & ? & ? & {\color{red} \CIRCLE} & ? & ? & 31 & (7\%) & 31 & (11\%) \\
? & {\color{red} \CIRCLE} & ? & {\color{red} \CIRCLE} & ? & ? & 29 & (6\%) & 28 & (10\%) \\
? & {\color{green} \CIRCLE} & ? & {\color{red} \CIRCLE} & ? & ? & 2 & (0\%) & 3 & (1\%) \\
? & ? & ? & {\color{red} \CIRCLE} & {\color{red} \CIRCLE} & ? & 7 & (1\%) & 7 & (2\%) \\
? & ? & ? & {\color{green} \CIRCLE} & {\color{red} \CIRCLE} & ? & 21 & (4\%) & 19 & (7\%) \\
{\color{red} \CIRCLE} & ? & ? & ? & ? & ? & 18 & (4\%) & 39 & (14\%) \\
{\color{red} \CIRCLE} & {\color{green} \CIRCLE} & ? & ? & ? & ? & 2 & (0\%) & 2 & (0\%) \\
{\color{red} \CIRCLE} & {\color{red} \CIRCLE} & ? & ? & ? & ? & 16 & (3\%) & 37 & (14\%) \\
{\color{green} \CIRCLE} & {\color{red} \CIRCLE} & ? & ? & ? & ? & 289 & (68\%) & 165 & (62\%) \\

\hline
\multicolumn{6}{c}{total} & 423 & & 263 &\label{output/data_table.tex}\unskip%
    \enddata
\end{deluxetable}

\begin{figure*}[h]
    \script{bg_core_prevs.py}
    \begin{centering}
        \includegraphics[width=\textwidth]{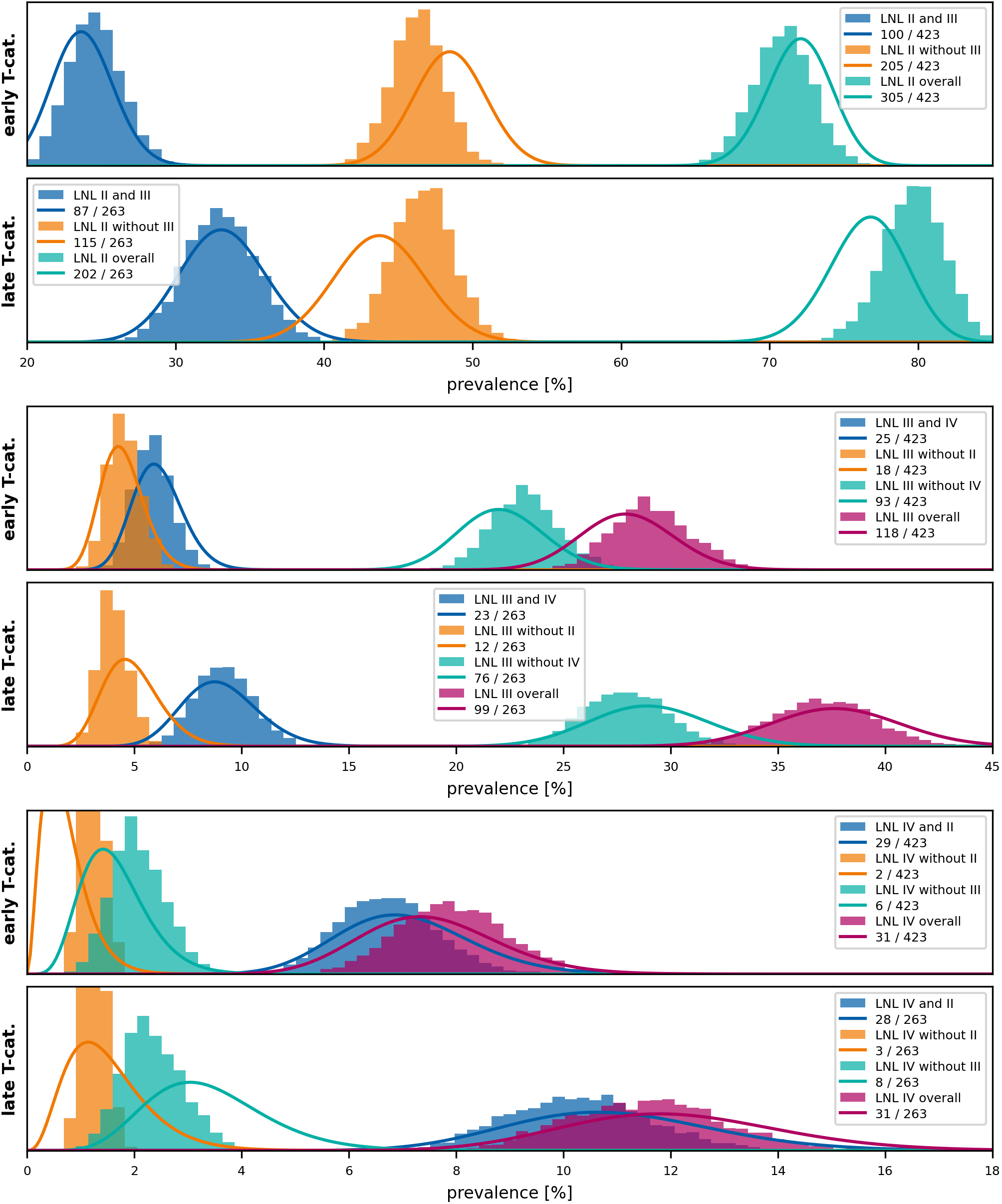}
        \caption{Prevalence of involvement as predicted by the base graph model for different scenarios involving the most commonly metastatic \glspl{lnl} II, III and IV (shaded histograms). The model's predictions are compared to Beta posteriors over the prevalence based on the frequency of the same scenarios and a uniform prior (slid lines). The top panel shows some selected scenarios with early T-category tumors and the bottom panel with advanced T-category. \label{fig:bg_prevalences}}
    \end{centering}
\end{figure*}

The process of computing the log-evidence using \gls{ti} was as follows: We randomly initialized the starting positions of the $W$ samplers in the ensemble within the $k$ dimensional unit cube $\boldsymbol{\Theta}$. Subsequently, for each $j \in \{ 0, 1, 2, \ldots, R=63 \}$ we drew samples from the corresponding power posterior with the value of $\beta_j$ set according to the power rule in \autoref{eq:power_rule}. This sampling at point $j$ consisted of 1000 burn-in steps, followed by 200 steps, of which only every tenth was kept. The last position of the $W$ chains for the $j$-th $\beta$ value in the ladder was used to initialize the subsequent sampling round with $\beta_{j+1}$. Hence, after the computations are finished, we are left with $S = 20 \cdot W$ samples $\boldsymbol{\hat{\theta}}_{i,j}$ and respective log-likelihood $\hat{\ell}_{i,j}$ from each of the 64 power posteriors corresponding to the respective $\beta_j$. Subsequently, we numerically integrated the following quantity $S$ times:
\begin{equation}
    \ln \hat{E}_i = \frac{1}{2} \sum_{j=0}^{R-1} \left( \beta_{j+1} - \beta_j \right) \cdot \left( \hat{\ell}_{i,j} + \hat{\ell}_{i,j+1} \right)
\end{equation}
And then computed the mean and standard deviation of all the integrated $\ln \hat{E}_i$. We then used this for the log-evidence and its error.

Without derivation or insight, we would like to mention that the model evidence naturally balances a model's accuracy against its complexity. The value of $\ln E$ will generally be larger (i.e., less negative) if a model fits the data better than another while being similarly complex. On the other hand, if e.g. additional parameters are introduced without sufficiently improving how well the model explains the data, the evidence will penalize the increase in complexity.

An approximation to the evidence that also attempts to balance accuracy and complexity against each other is the heuristic called \gls{bic}. The negative one half of the \gls{bic} approximates the $\ln E$ via Lagrange's method \citep{bhat_Derivation_2010} and yields an easy to compute estimate that may also be used to compare models, as long as its underlying assumptions are valid:
\begin{equation} \label{eq:bic}
    - \text{BIC} / 2 = \ln{\hat{\mathcal{L}}} - \frac{k}{2} \ln{N} \approx \ln{E}
\end{equation}
Here, $\hat{\mathcal{L}} = \max_{\boldsymbol{\theta}}{\left( \ln P \left( \boldsymbol{\mathcal{D}} \mid \boldsymbol{\theta} \right)\right)}$ is the maximum log-likelihood. The approximation is good, when the posterior distribution over the parameters $p\left( \boldsymbol{\theta} \mid \boldsymbol{\mathcal{D}} \right)$ is single-modal and falls quickly to zero from the maximum. Also, the number of data points $N$ needs to be much larger than the number of parameters $k$. We will see that for the models we consider here, the \gls{bic} is generally a good approximation and the conclusions drawn from comparing models using this metric can be reproduced reliably using the true model evidence computed with \gls{ti}.

\vspace{5mm}

\subsection{Reproducibility}
\label{subsec:complete_model:reproducibility}

The entire methodology used in this work is publicly available in the GitHub repository \href{https://github.com/rmnldwg/lynference}{\texttt{rmnldwg/lynference}}. Tagged references to specific versions of an inference pipeline allow reproducing the inferred parameters of the models described here. Every parameter necessary to reproduce such a pipeline is specified there in designated configuration files. It also defines the sequences of computations that constitute the pipeline that are mostly calls to commands of the \href{https://pypi.org/project/lyscripts/}{\texttt{lyscripts}} we published.

All figures in this work, like the risk predictions and prevalences we show histograms of, can be reproduced following the instructions of the GitHub repository underlying this publication \href{https://github.com/rmnldwg/graph-extension-paper}{\texttt{rmnldwg/graph-extension-paper}}. It is based off of the \href{https://show-your.work}{showyourwork} project that aims to make scientific papers easily and fully reproducible.

\section{Multicentric Dataset}
\label{sec:data}

The dataset $\boldsymbol{\mathcal{D}}$ that we used for inference is comprised of the detailed reports on lymph node involvement patterns in \gls{opscc} patients treated at three different institutions in France and Switzerland: The \gls{clb} in Lyon (France), the \gls{isb} in Bern (Switzerland), and the \gls{usz} in Zürich (Switzerland). We have previously published the patterns of nodal involvement for the \gls{usz} cohort (287 patients) \citep{ludwig_dataset_2022,ludwig_lydata_2023} and described its characteristics in detail \citep{ludwig_detailed_2022}. The first \gls{clb} dataset (263 patients) \citep{ludwig_lydata_2022} underlies a publication on \gls{hpv} status in \gls{opscc} \citep{bauwens_prevalence_2021} and is made available in a separate ``Data in Brief'' article alongside the second dataset from France \citep{ludwig_lydata_2023-2} and the lymphatic progression patterns from the \gls{isb} \citep{ludwig_lydata_2023-1,ludwig_multi-centric_2023}. All datasets may be explored online in our web-based interface \href{https://lyprox.org}{LyProX}.

In total, the dataset contains %
  686\label{output/num_patients.txt}\unskip%
patients with newly diagnosed \gls{opscc}. It includes patients treated with definitive (chemo)radiotherapy, adjuvant (chemo)radiotherapy following neck dissection, or neck dissection alone. Pathologically assessed \gls{lnl} involvement was available for 263 surgically treated patients, while for the remainder the nodal involvement was assessed based on available diagnostic modalities (FDG-PET-CT, CT, MRI, FNA). If multiple modalities were used to diagnose a patient's lymph node involvement, the available modalities were combined into a consensus decision. When different modalities were conflicting, the conflicts were resolved by inferring the most likely state (healthy or metastatic) for each \gls{lnl} separately. To do so, we used literature values for the sensitivity and specificity of the diagnostic modalities \citep{de_bondt_Detection_2007,kyzas_18f-fluorodeoxyglucose_2008}, which we also tabulated in \autoref{table:sens_spec}. Practically, this means that, whenever pathology after neck dissection was available, the pathology result was taken as the consensus, overruling any other clinical diagnostic modality. If, for example, PET-CT and MRI ware available and conflicting, PET-CT was taken as the consensus, overruling MRI.

The dataset containing the consensus decision for the involvement of each level in every patient was then used for model parameter learning. We assumed that it represents an observation of the true hidden state $\boldsymbol{\xi}_k$. The frequencies of some of the most important combinations of involved lymph node levels are listed in \autoref{table:data_prevalence}.

\begin{figure*}[!ht]
    \centering
    \includegraphics[width=\linewidth]{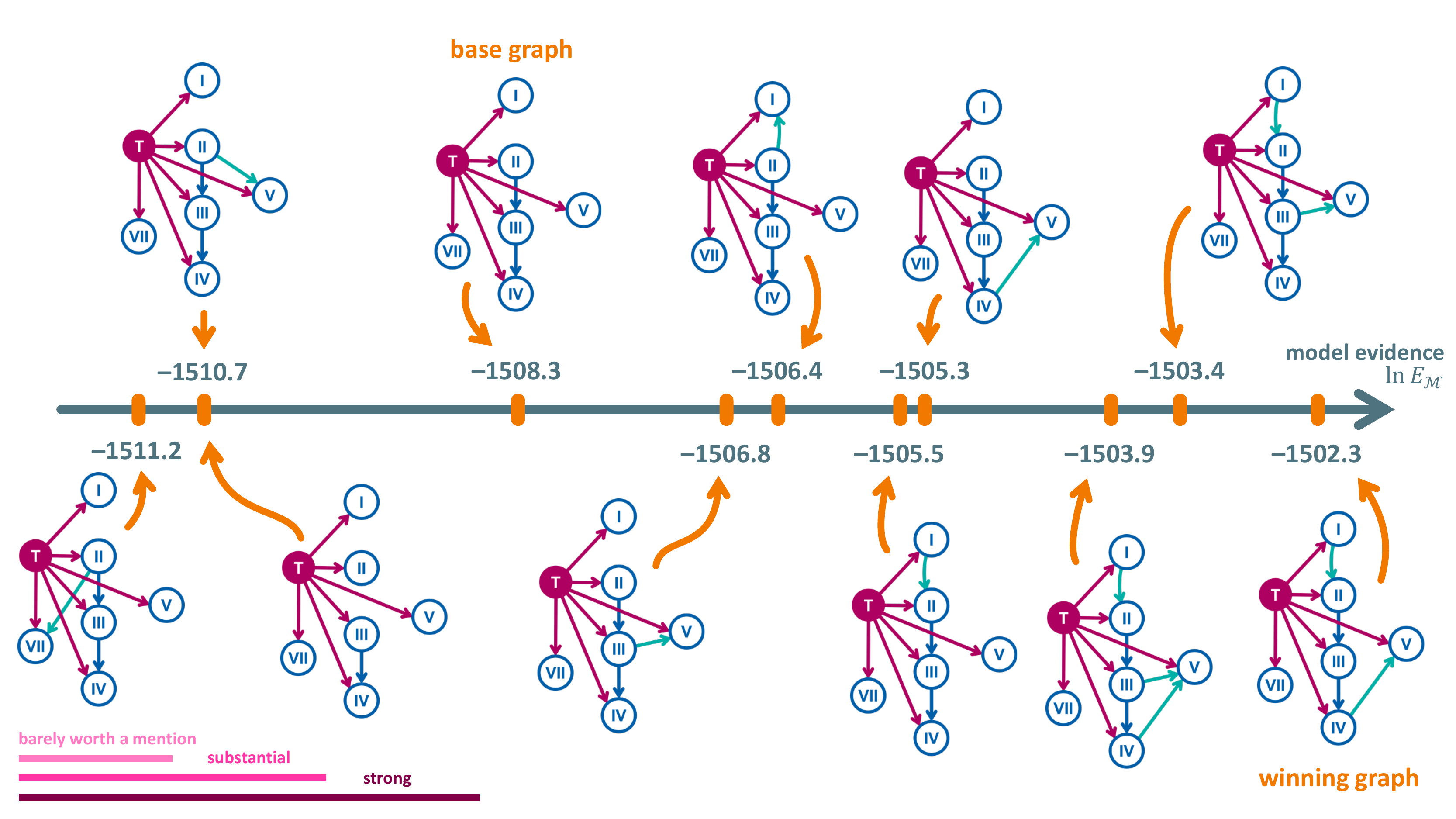}
    \caption{Visual ranking of the investigated graphs w.r.t. their model evidence, computed via thermodynamic integration. Not shown are the two graphs where the arc from \gls{lnl} III to IV was removed. Their respective model evidence fell below $-1528$ and the two graphs would appear far left in the figure. In the bottom left corner, we provide a visual reference in analogy to \autoref{table:bayes_factor}: E.g., any difference in the model evidence shorter than the first of the three rulers indicates that the improvement is ``barely worth a mention''. \label{fig:ranking}}
\end{figure*}

\section{Results}
\label{sec:results}

\subsection{Involvement of Levels II, III, and IV}
\label{subsec:results:lnls_II_III_IV}

For the base graph, we have plotted the predicted prevalence of involvement patterns in the investigated patient cohort for scenarios involving the most commonly metastatic \glspl{lnl} II, III and IV in \autoref{fig:bg_prevalences}. It shows -- for each pattern of lymphatic involvement -- two plots overlaid:

\begin{enumerate}
    \item The colored histograms over the base graph model's prediction for the prevalence of the respective pattern of involvement. These histograms are obtained by computing the same prevalence with different samples from the inference process, thus providing us with a measure of uncertainty for the prediction.
    \item Colored lines, depicting the beta posterior over the same involvement pattern's prevalence, given a uniform beta prior and the binomial likelihood of the observed data. The maximum of the beta distributions always coincides with the data prevalence but we additionally gain an intuition into how statistically significant the data is. E.g., Observing 3 out of 10 patients with a particular pattern of nodal metastases is less convincing than 300 out of a cohort of 1000 patients. A beta posterior over these prevalences reflects that in its variance.
\end{enumerate}

\autoref{fig:bg_prevalences} shows that this minimal graph is already capable of describing the most important parts of the observed data very well. Notably, the model is not only accurate in its predictions, it also correctly estimates the variance stemming from the limited amount of data. The separation between involvement prevalences of early and advanced T-category tumors is also reproduced well by the model. This is remarkable because the model introduces only a single parameter to describe the differences between early and advanced T-category for all involvement patterns. This shows that expecting later diagnosis times, on average, for patients with advanced T-category tumors can explain more severe lymphatic involvement.

It is interesting to note that not all involvement patterns become more prevalent with advanced T-category. For example, a healthy \gls{lnl} III together with a metastatic level II is observed slightly less often for advanced T-category tumors compared to early T-category (yellow histogram, row 1 versus 2). This is because, for advanced T-category, it is more likely the disease has already spread to \gls{lnl} III (blue histogram, row 1 versus 2). Our model captures this accurately and precisely.

\subsection{Comparison of candidates graphs}
\label{subsec:results:graph_candidates}

\begin{deluxetable}{lr}[t]
    \tablecaption{Model comparison results from the base graph and the extended graph we chose as the ``winning'' model. For both \glspl{dag} we show the log-evidence, computed via thermodynamic integration, the negative one half of the \gls{bic}, as well as the maximum log-likelihood that was encountered during the final \gls{mcmc} sampling round. \label{table:evidence}}
    \tablehead{
        \textbf{graph}  & \textbf{log-evidence}
    }
    \startdata
  add I → II \& IV → V & -1502.33  \\ 
add I → II \& III → V & -1503.36  \\ 
add I → II \& III → V \& IV → V & -1503.88  \\ 
add IV → V & -1505.28  \\ 
add I → II & -1505.46  \\ 
add II → I & -1506.37  \\ 
add III → V & -1506.76  \\ 
base graph & -1508.32  \\ 
remove II → III & -1510.67  \\ 
add II → V & -1510.68  \\ 
add II → VII & -1511.17  \\ 
remove III → IV & -1528.71  \\ 
remove II → III \& III → IV & -1530.79 
\label{output/metrics_table.tex}\unskip%
    \enddata
\end{deluxetable}

The model evidences of all candidate graphs are reported in \autoref{table:evidence}. A visual ranking is provided in \autoref{fig:ranking}. Let us first consider the six models in which one of the candidate arcs is added to the base graph. Evidently, adding a connection from \gls{lnl} I to II is strongly supported given this dataset, and is slightly superior to the reverse connection from \gls{lnl} II to I. In addition, there is strong evidence for introducing an arc from \gls{lnl} IV to V. Furthermore, there is substantial evidence for an arc from \gls{lnl} III to V. All other investigated additions lead to improvements that are barely worth a mention or do not justify the additional complexity at all, indicated by a lower evidence. 

Based on these results, we consider three additional candidates for the optimal graph that combine the added arc from \gls{lnl} I to II with the arc(s) $\text{III} \rightarrow \text{V}$ and/or $\text{IV} \rightarrow \text{V}$. The model evidence for these three graphs is also reported in \autoref{table:evidence}. The best performing graph with decisive evidence over the base graph turned out to be the one which combines the arcs from \gls{lnl} I to II and IV to V. Interestingly, the evidence gain of this ``winning graph'' is roughly the sum of the gains seen in the two candidates where only one of these connections was added, respectively. This indicates that the two additional parameters are largely independent of each other and manage to describe different aspects of the data.

\begin{figure*}
    \script{comp_IIIandIVandV_prevs.py}
    \begin{centering}
        \includegraphics[width=\linewidth]{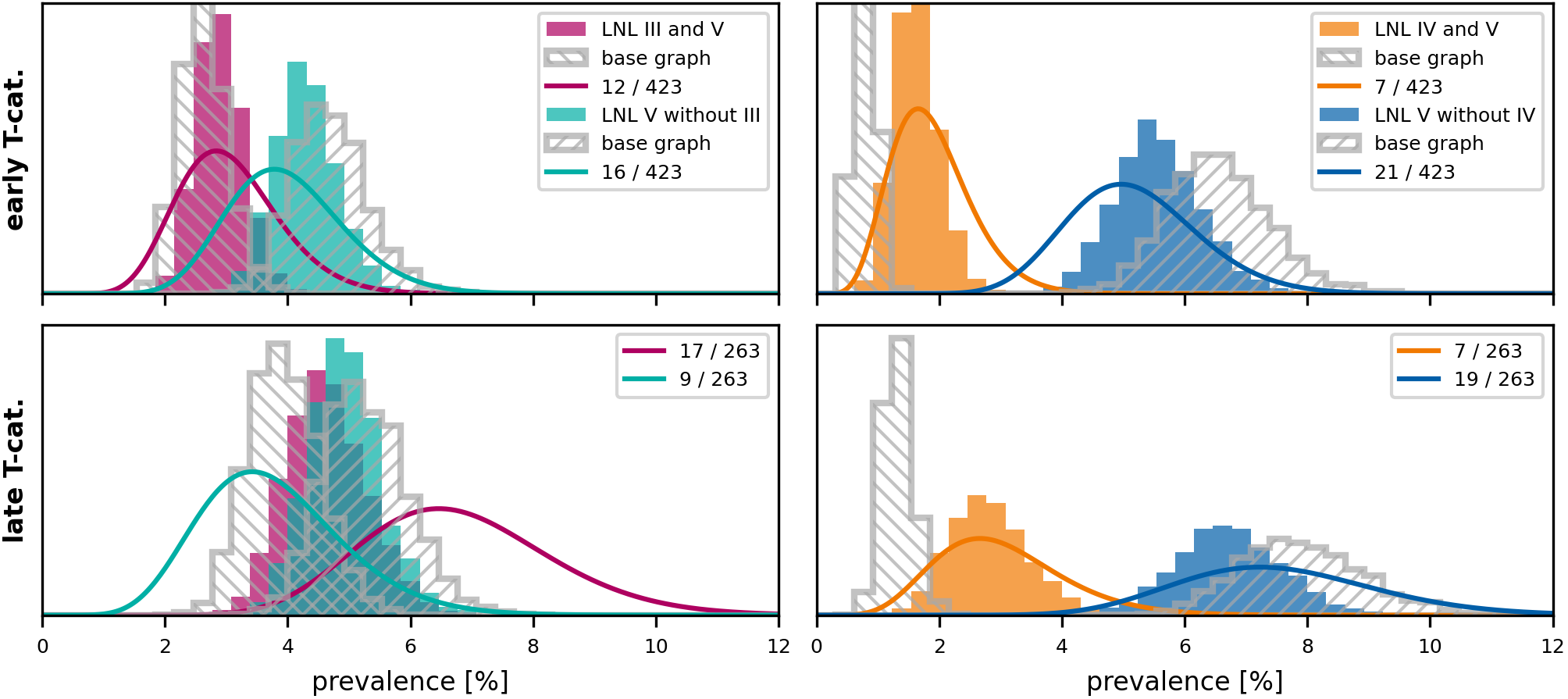}
        \caption{Observed (Beta posteriors as lines) vs. predicted (histograms) prevalences of involvement combinations that include \gls{lnl} V. We have plotted the predictions from the winning graph (colored histograms) and those of the base graph (gray, hatched histograms). The top two panels show scenarios for early T-category patients, the bottom two panels for advanced T-category. The left two panels consider combinations of \gls{lnl} III and V involvement, while the right two panels consider combinations of \gls{lnl} IV and V. The colored lines show the Beta posterior over the prevalence of the respective involvement pattern, given the data. \label{fig:IIIandIVandV_prevs}}
    \end{centering}
\end{figure*}

\autoref{table:evidence} additionally shows the evidence of graphs in which the arc from level II to III or from level III to IV is removed. The low model evidence for these graphs confirms the importance of these connections and is consistent with the anatomical motivation. The connection from level III to IV is crucial for describing the observation that metastases in level IV are extremely rare without simultaneous involvement of the upstream level III. \looseness=-1

\subsection{The winning graph}
\label{subsec:results:optimal_graph}

The most likely model parameters for the winning graph, corresponding to the mean of the marginals of the sampled posterior distributions, are tabulated in \autoref{table:params}. We have fixed $p_\text{early}=0.3$ for early T-category tumors (i.e., T0, T1, and T2), and $t_\text{max}=10$ time steps. The result that $t_{\text{II} \rightarrow \text{III}}$ and $t_{\text{III} \rightarrow \text{IV}}$ are relatively large compared to $b_\text{III}$ and $b_\text{IV}$ reflects the observation that skip metastases in levels III and IV without involvement of the upstream level are rare. Since level II's parent node (level I) is rarely involved, $b_\text{II}$ can approximately be related to the prevalence of involvement in level II. The probability for no involvement of level II when the patient is diagnosed after $t$ time steps is $\left(1-b_\text{II}\right)^t$. The prevalence of level II involvement for advanced T-category patients is thus
\begin{equation}
    \begin{aligned}
        \text{prev}_\text{late}^\text{II} &= 1-\sum_{t=0}^{10} \left(1-b_\text{II}\right)^t \cdot p_\text{late}^t \, (1-p_\text{late})^{(10-t)}\binom{10}{t} \\
        &\approx 79 \%
    \end{aligned}
\end{equation}

which agrees with the second panel from the top in \autoref{fig:bg_prevalences}. 
The large value for the parameter $t_{\text{I} \rightarrow \text{II}}$ reflects the observation that in almost all patients with level I involvement, level II is also involved. The large uncertainty in $t_{\text{I} \rightarrow \text{II}}$ is related to the fact that level I involvement is rare compared to level II.

\subsection{Involvement of levels I and V}

We can observe that the winning graph describes the involvement of levels II, III, and IV equally well as the base graph, a result that is expected and not further shown. We thus focus on the improvements w.r.t. involvement patterns that include the \glspl{lnl} I and V, that more rarely harbor metastases.

\begin{figure*}
    \script{comp_IandII_prevs.py}
    \begin{centering}
        \includegraphics[width=\linewidth]{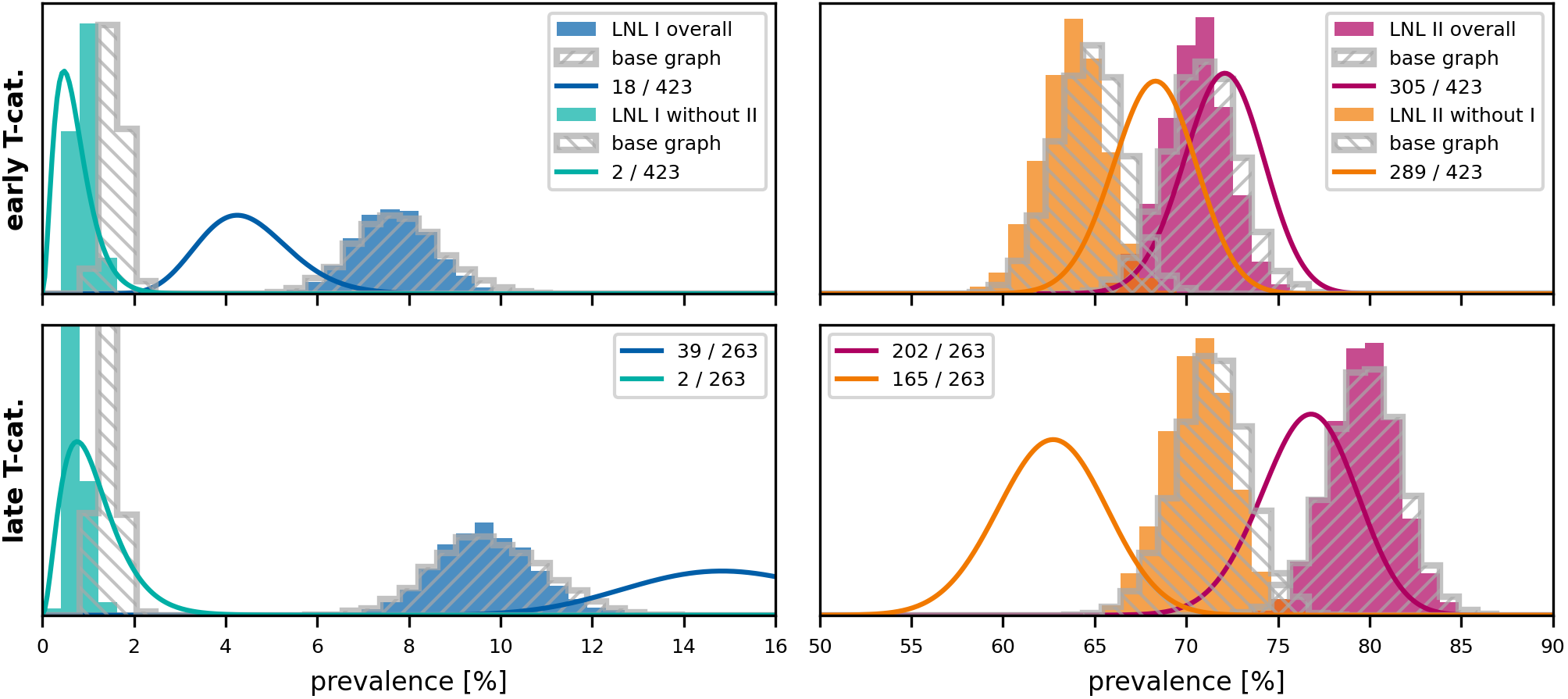}
        \caption{Comparison of observed and predicted prevalences of \gls{lnl} I and II involvement patterns. The top and bottom panels show the prevalences for early and advanced T-category, respectively. The solid lines are Beta posteriors from the data, while the histograms are predicted prevalences (colored: winning graph, gray-hatched: base graph). Blue and red plots indicate overall \gls{lnl} I and II involvement, respectively. Green plots indicate \gls{lnl} I involvement without level II, while orange plots indicate the opposite (\gls{lnl} II without level I). The winning graph has an added edge from \gls{lnl} I to II, which improves the prediction of the rare green pattern. \label{fig:IandII_prevs}}
    \end{centering}
\end{figure*}

{\it Level V:} In \autoref{fig:IIIandIVandV_prevs} we compare the base graph's and the winning graph's estimations for prevalences of involvement patterns that include \gls{lnl} V. The base graph underestimates the probability that level IV and V are simultaneously involved, and overestimates the probability that level V but not IV is involved. By introducing the arc from level IV to V, the winning graph can describe the observation that level V involvement is typically associated with severe involvement of level II-IV. In the dataset, 14 patients out of 62 patients with level IV involvement have metastases in level V (22 \%). Instead, only 40 patients out of 624 patients without level IV involvement have metastases in level V (6 \%).

{\it Level I:} In \autoref{fig:IandII_prevs}, analogous comparisons are shown for involvement patterns that include \gls{lnl} I. The base graph overestimates the probability of level I involvement without simultaneous involvement of level II. By introducing the arc from level I to II, the winning graph can capture the correlations between levels I and II. It can also be noted that both models overestimate level I involvement for early T-category patients and underestimate level I involvement for advanced T-category patients. This is further described in the discussion section below.

\begin{deluxetable}{lrr}[b]
    \tablecaption{Mean and standard deviation of parameters sampled for the winning graph in percent. \label{table:params}}
    \tablehead{
        \textbf{parameter} & \textbf{mean} & \textbf{std. dev.}
    }
    \startdata
  $b_\text{I}$ & 2.65 \% & $\pm$ 0.31 \%  \\ 
$b_\text{II}$ & 37.67 \% & $\pm$ 1.81 \%  \\ 
$b_\text{III}$ & 8.1 \% & $\pm$ 1.26 \%  \\ 
$b_\text{IV}$ & 1.1 \% & $\pm$ 0.24 \%  \\ 
$b_\text{V}$ & 2.13 \% & $\pm$ 0.28 \%  \\ 
$b_\text{VII}$ & 2.16 \% & $\pm$ 0.31 \%  \\ 
$t_{\text{I} \rightarrow {II}}$ & 66.76 \% & $\pm$ 21.37 \%  \\ 
$t_{\text{II} \rightarrow {III}}$ & 9.49 \% & $\pm$ 3.04 \%  \\ 
$t_{\text{III} \rightarrow {IV}}$ & 14.48 \% & $\pm$ 2.43 \%  \\ 
$t_{\text{IV} \rightarrow {V}}$ & 14.57 \% & $\pm$ 5.29 \%  \\ 
$p_\text{late}$ & 38.34 \% & $\pm$ 2.26 \% 
\label{output/means_table.tex}\unskip%
    \enddata
\end{deluxetable}

\subsection{Risk Prediction for Occult Disease}
\label{subsec:results:risk_prediction}

In this section, the model corresponding to the winning graph is applied to estimating the risk of occult metastases in clinically negative \glspl{lnl}. We assume a sensitivity of 0.76 and a specificity of 0.81 for the clinical diagnosis of lymph node metastases, corresponding to CT imaging in \autoref{table:sens_spec}.

{\it Level II:} As can be seen in \autoref{table:params}, spread from the tumor to \gls{lnl} II to be the most probable transition at any given time step. As a consequence, even for an early T-category patient that presents with a clinical N0 neck, our model predicts a %
  31.13 \% $\pm$ 1.78 \%\label{output/wg_II_risks.tex}\unskip%
risk for microscopic metastases in \gls{lnl} II.

\begin{figure}[b]
    \script{wg_III_risks.py}
    \begin{centering}
        \includegraphics[width=\linewidth]{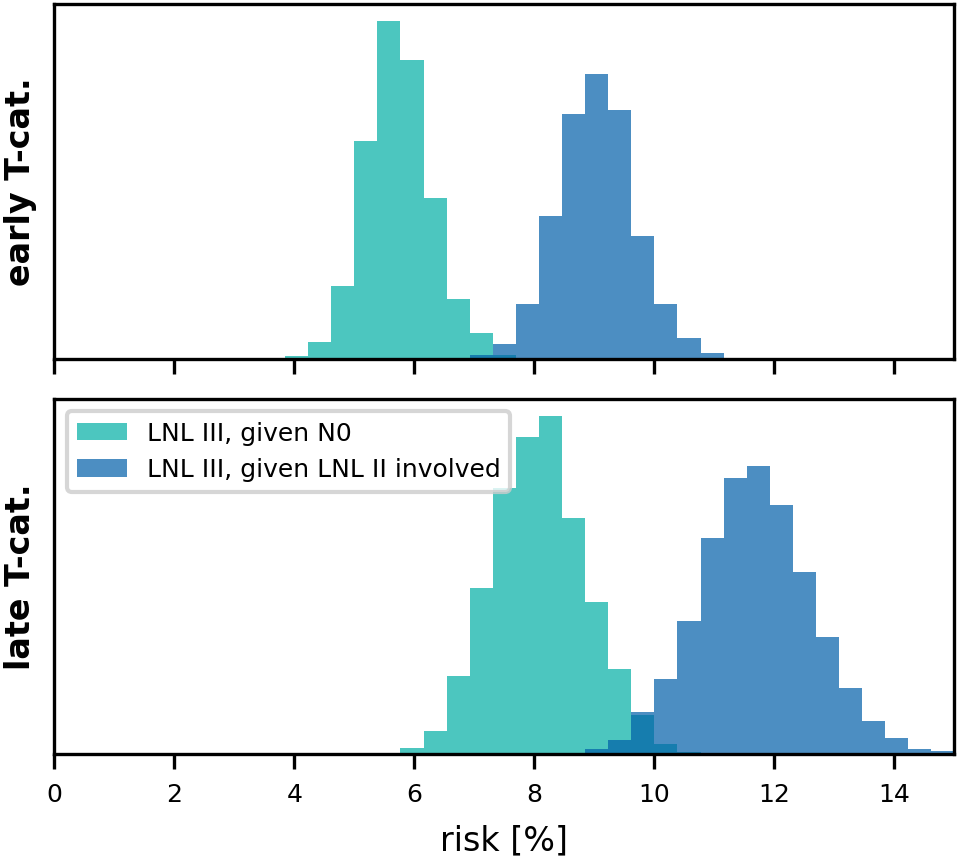}
        \caption{Histograms over the risk for microscopic involvement in \gls{lnl} III, given that a patient presents as clinically N0 (green), or given that \gls{lnl} II shows clinical involvement (blue). The top panel shows these risks for early T-category, the bottom panel for late. \label{fig:wg_III_risks}}
    \end{centering}
\end{figure}

\begin{figure*}
    \script{wg_IVandV_risks.py}
    \begin{centering}
        \includegraphics[width=\linewidth]{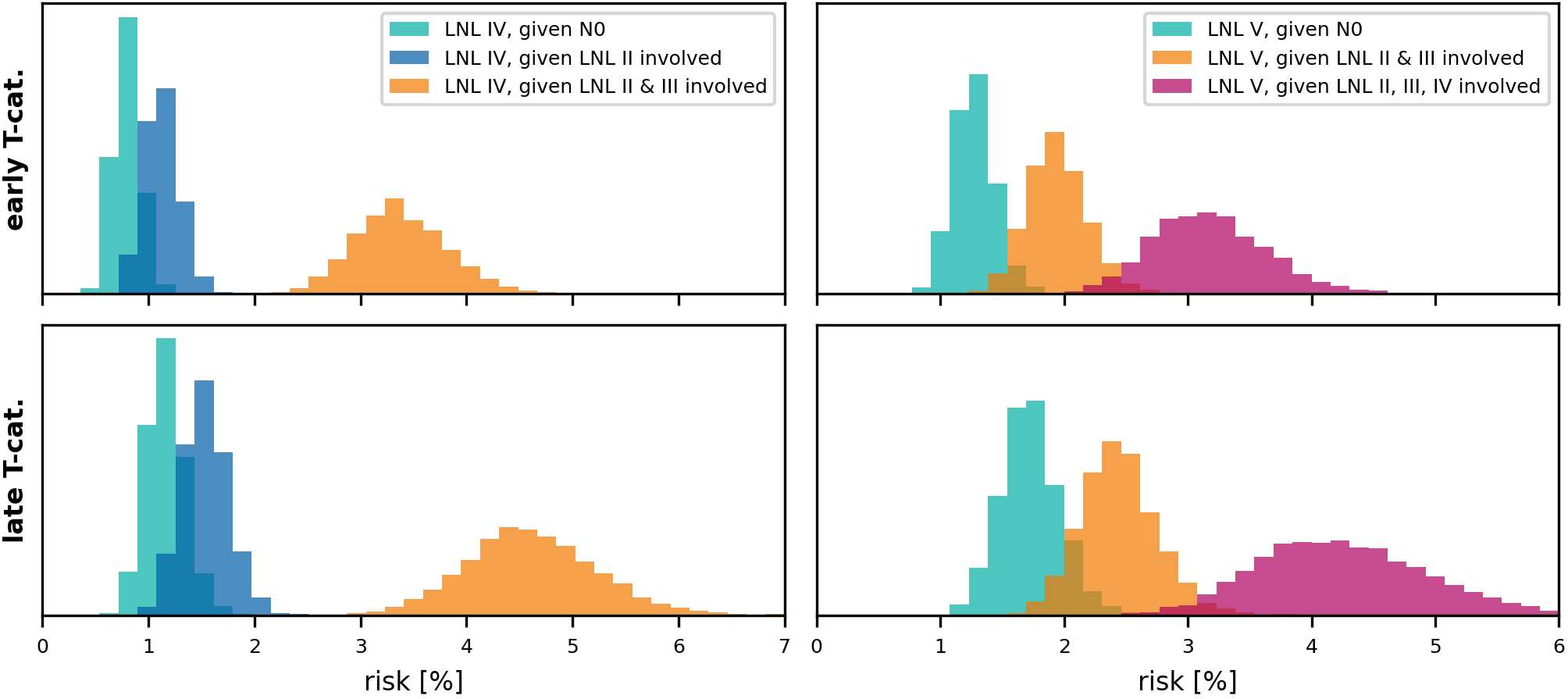}
        \caption{Distributions over the risk for microscopic involvement in \gls{lnl} IV (left panels) and in \gls{lnl} V (right panels) as predicted by the winning graph model, given early (top panels) or advanced T-category (bottom panels), and different CT-based diagnoses: 1) A clinical N0 patient (green histograms), 2) visible metastases in \gls{lnl} II, but otherwise healthy-looking lymph nodes (blue histograms), 3) macroscopic metastases in the \glspl{lnl} II \& III, with the rest of the neck still being clinically node negative (orange histograms), and finally 4) extensive clinical involvement in the levels II, III, and IV (red histograms).\label{fig:wg_IVandV_risks}}
    \end{centering}
\end{figure*}

{\it Level III:} \autoref{fig:wg_III_risks} compares the risk of occult disease in level III between patients that are clinically N0 (orange) and patients with clinically diagnosed involvement of only level II (red), for early T-category (upper panel) and advanced T-category (bottom panel). The histograms represent the uncertainty in the model's risk prediction arising from the uncertainty in the model parameters and are generated by randomly drawing a tenth of the samples from the model parameter's joint posterior distribution. This amounts to $S = 20 \cdot W$ samples, as described in \autoref{subsec:complete_model:comparison}. The model predicts a risk of just below 6\% for early T-category tumors and 8\% for T3 or T4 ones. For patients with involvement of level II, the risk in level III increases to approximately 9\% and 12\%, respectively.

{\it Level IV:} \autoref{fig:wg_IVandV_risks} (left panels) compares the risk of occult disease in level IV for the typical clinical presentations: clinically N0 (green), metastases in level II (blue), and metastases in levels II and III (orange). The model predicts a low risk of 1-2\% in level IV for patients with clinically healthy level III. For patients with clinical involvement of level III, the risk of occult disease in level IV increases to approximately 3\% for early T-category and 5\% for advanced T-category tumors.

{\it Level V:} The right panels in \autoref{fig:wg_IVandV_risks} show the risk of occult disease in level V depending on T-category and the clinical involvement of levels II-IV. For clinically N0 patients, the risk in level V is estimated to be just above 1\%. Extensive nodal involvement of levels II-IV increases the risk in level V to more than 4\% for advanced T-category tumors. 

\begin{figure}
    \script{wg_I_risks.py}
    \begin{centering}
        \includegraphics[width=\linewidth]{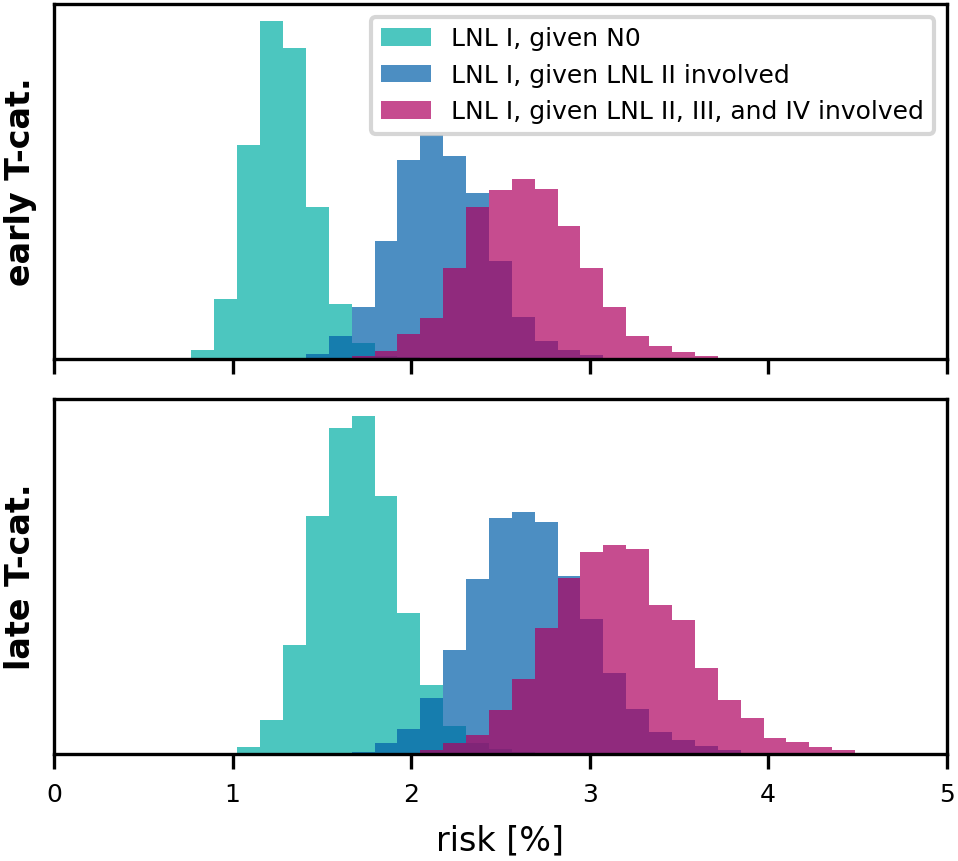}
        \caption{Distributions over predicted risk for involvement in \gls{lnl} I, given different clinical diagnosis scenarios: For N0 patients (green), patients with macroscopic involvement in \gls{lnl} II (blue), and for the case where the \glspl{lnl} II, III, and IV show involvement. The top panel shows these risks for early T-category and the bottom row for advanced T-category. \label{fig:wg_I_risks}}
    \end{centering}
\end{figure}

{\it Level I:} \autoref{fig:wg_I_risks} shows the risk of occult disease in level I depending on T-category and the clinical involvement of levels II-IV. For clinically N0 patients, the risk in level I is estimated to be in the order of 1-2\%. Extensive nodal involvement of levels II-IV increases the risk in level I to just below 4\% for advanced T-category tumors. It is pointed out that the winning graph does not contain arcs from levels III or IV to \gls{lnl} I (and anatomically we do not assume that there is lymphatic drainage from levels III or IV to level I). Thus, the increased risk in level I is related to the time evolution: Getting diagnosed at a later time during the disease's evolution probably correlates with more advanced nodal metastasis. And involvement in the levels III and IV corresponds to a more advances state of disease that is likely diagnosed at a later time step, such that the tumor also had more time to spread to level I. The correlation between the clinical involvement pattern, the likely time of diagnosis in the tumor's time frame based on it, and from that the risk of involvement is another benefit of the formulation of the model as a \gls{hmm}.

\begin{figure}[b]
    \script{wg_any_risks.py}
    \begin{centering}
        \includegraphics[width=\linewidth]{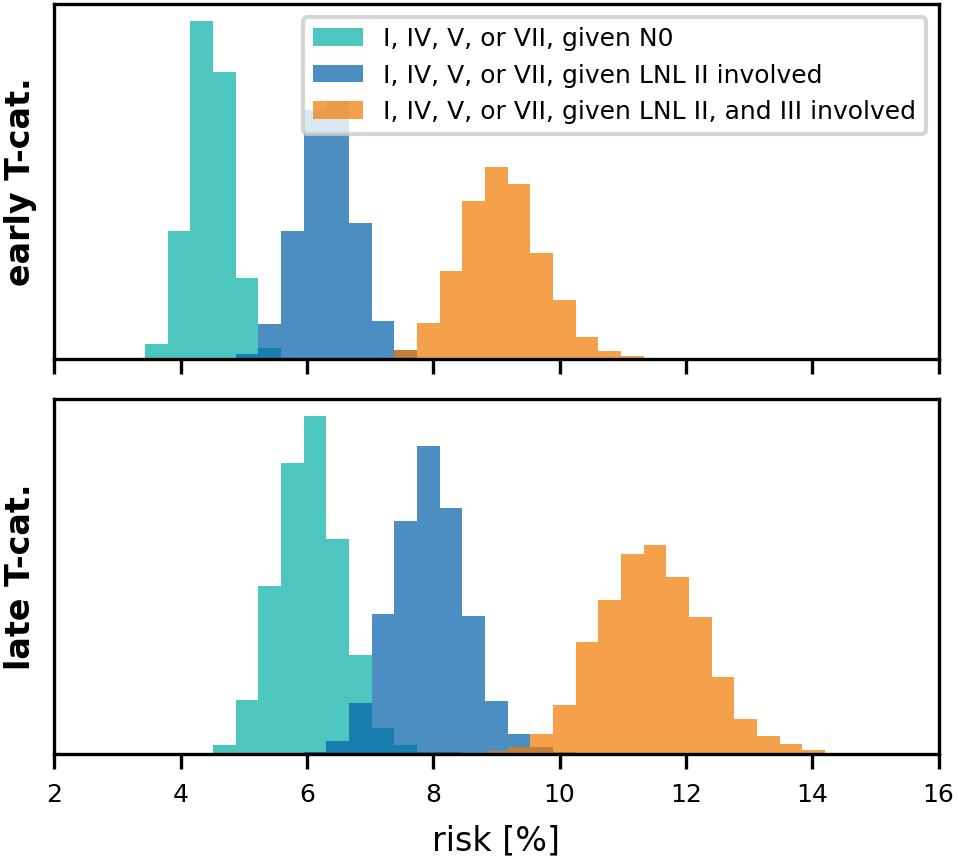}
        \caption{Shown are the histograms over the predicted risk for involvement in any of the \glspl{lnl} I, IV, V, or VII. The risk is plotted given a clinical N0 diagnosis (green), macroscopically detected metastases in \gls{lnl} II (blue), and lastly given visible involvement in both \gls{lnl} II and III (orange). The top row shows these risks for early T-category diagnoses and the bottom row for advanced T-category. \label{fig:wg_any_risks}}
    \end{centering}
\end{figure}

To illustrate the flexibility of the model in predicting various risks, we have plotted the risk for occult disease in \textit{any} of the \glspl{lnl} I, IV, and/or V, given different clinical diagnoses in \autoref{fig:wg_any_risks}. Similar to this, we may compute the risk for an arbitrary combination of involved levels, given a similarly arbitrary clinical diagnosis. For the base graph (\href{https://lyprox.org/riskpredictor/9}{\texttt{base-graph-v2}}) and the winning graph (\href{https://lyprox.org/riskpredictor/8}{\texttt{win-graph-v3}}), one may also interactively explore these risks in our web-based interface \href{https://lyprox.org/riskpredictor/list}{LyProX}, similar to how it is possible to explore the underlying data in an interactive way.

\clearpage

\section{Discussion}
\label{sec:discussion}

\subsection{Summary} 

In this publication we present a statistical model of ipsilateral lymph node involvement in oropharyngeal SCC patients. Although the basic \gls{hmm} of lymphatic progression has been conceptually introduced in a previous work \citep{ludwig_hidden_2021}, this is the first publication that evaluates the model based on a large multi-institutional dataset containing %
  686\label{output/num_patients.txt}\unskip%
patients. It is demonstrated for the first time that the model can accurately describe the patterns of lymph node involvement observed in the data, including the correlations between levels and its dependence on T-category. Furthermore, techniques from statistical physics are applied to calculate the model evidence for Bayesian model comparison. This yields a complete model including all \glspl{lnl} relevant for \gls{opscc}: I, II, III, IV, V, and VII with a parameterization that balances accuracy and model complexity. 

\subsection{Implications for elective nodal treatment} 
Risk predictions obtained by the model may be used to design clinical trials on volume-deescalated treatment of \gls{opscc}. In the context of radiotherapy, this corresponds to excluding \glspl{lnl} from the \gls{ctv-n}, which are irradiated according to the current guidelines. The list below should be seen as a summary of \autoref{subsec:results:risk_prediction} and the limitations discussed in \autoref{subsec:disc:limitation} should be taken into account in its interpretation. Assuming that one accepts approximately a 5\% risk of occult metastases per \gls{lnl}, the statistical model presented in this paper would suggest to:
\begin{itemize}
    \item[$\bullet$] Irradiate level II for all patients.
    \item[$\bullet$] Irradiate level III for most patients. Only for clinically N0, early T-category patients, not irradiating level III could be considered. 
    \item[$\bullet$] Exclude level IV from the \gls{ctv-n} for patients with clinically negative level III. For advanced T-category patients with involvement of level III, level IV should be irradiated.
    \item[$\bullet$] Exclude level V from the \gls{ctv-n} for most patients. Only for patients with extensive involvement of levels II, III, and IV, irradiation of level V can be considered.
    \item[$\bullet$] Exclude level I from the \gls{ctv-n} for early T-category patients with limited metastatic disease. For advanced T-category patients with extensive nodal involvement of levels II, III, and IV, level I should be irradiated (see also the limitations discussed in \autoref{subsec:disc:limitation}).
    \item[$\bullet$] Exclude level VII in all patients, unless it is clinically involved.
\end{itemize}

\subsection{Limitations and future work}
\label{subsec:disc:limitation}

{\it T-category dependence of level I involvement:} As shown in \autoref{fig:bg_prevalences}, the model describes the involvement of \glspl{lnl} II, III, and IV depending on T-category very well despite having only a single parameter related to T-category. \autoref{fig:IandII_prevs} shows that the model does not perfectly describe the T-category dependence of level I. It adjusts the parameters such that level I involvement is correctly described for the set of all patients combined, but it overestimates level I involvement for early T-category and underestimates it for advanced T-category. A possible explanation is that advanced T-category tumors are more likely to have grown into regions with direct lymph drainage to level I. The more severe involvement in levels II, III, IV for advanced T-category can be explained by tumors having more time to spread while keeping the spread probability rates $b_2$, $b_3$, $b_4$ constant. Regarding level I, early versus advanced T-category tumors the model may need different spread probability rates $b_1$ to describe their involvement correctly.

\begin{figure}[t]
    \script{wg_sens_spec_risks.py}
    \begin{centering}
        \includegraphics[width=\linewidth]{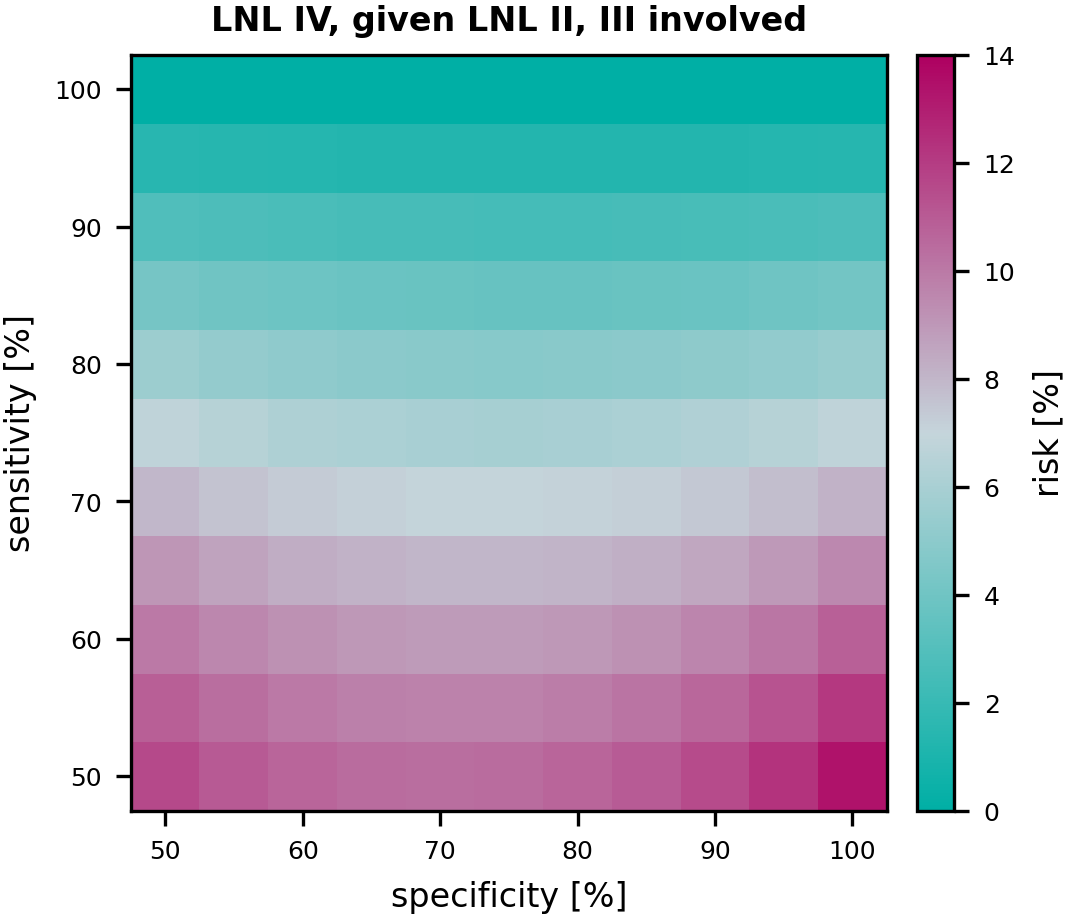}
        \caption{Dependence of risk of occult disease on sensitivity and specificity. The figure shows the risk of involvement in level IV, given the mean of the parameter samples, drawn during sampling, in patients with clinical involvement of levels II and III. \label{fig:wg_sens_spec_risks}}
    \end{centering}
\end{figure}

{\it Sensitivity and Specificity:} Estimating the risk of occult metastases depends on the assumed parameter values for sensitivity and specificity of clinical detection of lymph node metastases. For this work, we adopted literature values for sensitivity and specificity. However, different authors have estimated these values using different criteria and different methods. Consequently, these values need to be considered with caution. \autoref{fig:wg_sens_spec_risks} illustrates for one example how the risk of occult disease depends on sensitivity and specificity. Here, we consider the risk in level IV in patients with clinically detected metastases in levels II and III. For our default parameters of 81\% specificity and 76 \% sensitivity, the risk is 5\%, but it increases to around 8\% for a a sensitivity of 66\%.

Also, as described in \autoref{sec:data}, we assumed the consensus of the data to represent the true state of nodal involvement. This was not strictly necessary: Instead of computing a consensus beforehand and providing that with sensitivity and specificity of 1 to the model, as if it were the ground truth, we could have provided multiple diagnostic modalities per patient to the model directly. In fact, for patients with a pathology report available, this would even yield the same results. But we also decided to consider the consensus as an observation of the true hidden state for patients without pathologically assessed involvement. We did this because the literature values for sensitivity and specificity of around 80\% do not plausibly match the observation that around 78\% of patients in the \gls{usz} cohort showed clinical \gls{lnl} II involvement. The most likely true prevalence of involvement in \gls{lnl} II would need to be close to 100\%.

Discussing the possible origins for this discrepancy is beyond the scope of this work. Assuming the consensus to represent the true hidden state of a patient nonetheless allowed us to investigate if the model can describe plausible patterns of nodal involvement well. Future work may aim at developing new methods to model the difference between pathological and clinical lymph node involvement based on surgically treated patients in whom both is reported.

\begin{acknowledgments}

This work was supported by the Swiss Cancer Research Foundation under grant \href{https://www.krebsforschung.ch/unterstuetzen-sie-uns/stiftungen/-dl-/fileadmin/downloads/unterstuetzen-sie-uns/projekte-der-stiftung-krebsforschung-schweiz-2023.pdf}{KFS 5645-08-2022} and by the University Zürich under the Clinical Research Priority Program \href{https://www.crpp-ai-oncology.uzh.ch/en/Projects/Project-5.html}{Artificial Intelligence in Oncological Imaging}.

\end{acknowledgments}

\clearpage

\bibliography{bib}

\end{document}